\documentstyle[12pt]{article}

\newcounter{eq}
\newcounter{sc}

\def\overleftrightarrow#1{\vbox{\ialign{##\crcr
 $\leftrightarrow$\crcr\noalign{\kern-1pt\nointerlineskip}
 $\hfil\displaystyle{#1}\hfil$\crcr}}}

\def\slash#1{\not\!#1}

\newlength{\minitwocolumn}
\begin{document}

\begin{flushright}
DPUR/TH/39\\
August, 2013\\
\end{flushright}
\vspace{20pt}

\pagestyle{empty}
\baselineskip15pt

\begin{center}
{\large\bf Higgs Mechanism in Scale-Invariant Gravity
\vskip 1mm }

\vspace{20mm}
Ichiro Oda \footnote{E-mail address:\ ioda@phys.u-ryukyu.ac.jp
}

\vspace{5mm}
           Department of Physics, Faculty of Science, University of the 
           Ryukyus,\\
           Nishihara, Okinawa 903-0213, Japan.\\

\end{center}


\vspace{5mm}
\begin{abstract}
We consider a Higgs mechanism in scale-invariant theories of gravitation. 
It is shown that in spontaneous symmetry breakdown of scale invariance, gauge symmetries
are also broken spontaneously even without the Higgs potential if the corresponding charged scalar
fields couple to a scalar curvature in a non-minimal way. In this gravity-inspired new Higgs mechanism, 
the non-minimal coupling term, which is scale-invariant, plays a critical role. 
Various generalizations of this mechanism are possible and particularly the generalizations to non-abelian gauge 
groups and a scalar field with multi-components are presented in some detail. Moreover, we apply our finding
to a scale-invariant extension of the standard model (SM) and calculate radiative corrections. In particular,
we elucidate the coupling between the dilaton and the Higgs particle and show that the dilaton mass takes a value 
around the GeV scale owing to quantum effects even if the dilaton is massless at the classical level.  
\end{abstract}

\newpage
\pagestyle{plain}
\pagenumbering{arabic}


\rm
\section{Introduction}
Current understanding of elementary particle physics is based on two celebrated fundamental
principles, which are gauge symmetry and spontaneous symmetry breakdown of the gauge symmetry.   
In four interactions among elementary particles, strong, weak and electro-magnetic interactions are known to be
described on the same footing in terms of a gauge theory which is the standard model (SM) on the basis
of $SU(3) \times SU(2) \times U(1)$ gauge group, and gravitational interaction is believed to be also
described by a gauge theory whose final formalism is still far from complete at present.

The gauge principle alone, however, cannot describe the known structure of elementary particles.
The gauge principle requires elementary particles to be massless\footnote{To tell this statement more 
precisely, gauge invariance with the conventional form forbids the presence of a mass term of gauge
field, but such a mass term can exist if we change the gauge transformation to a more complicated 
form. This situation happens when we consider a theory with spontaneous symmetry beakdown where
a translation of the field by a constant must be accompanied \cite{Iliopoulos}.},  so in order to generate masses
for elementary particles the $SU(2) \times U(1)$ gauge symmetry must be spontaneously broken
at any rate. The idea of spontaneous symmetry breakdown itself is not new for elementary particle
physics but has emerged as a universal phenomenon in physics, in particular, condensed matter
physics. An alternative and indeed older description of super-conductivity, which was developed 
by Ginzburg and Landau, turned out to be a phenomenological representation of the BCS theory
\cite{Ginzburg}.  In this transcription, the complex "Ginzburg-Landau" scalar field is nothing but
the Higgs boson representing a bound pair of electrons and holes, and its phase and amplitude
components correspond to the massless Nambu-Goldstone boson and the massive Higgs type of 
excitations, respectively. As it happens, the collective excitations of both the types do exist
in all phenomena of the super-fluidity type. This universality of the spontaneous symmetry breakdown,
however, seems to have no implication in gravity so far. The concept of the mass is intimately connected
with general relativity since the right-hand side of Einstein's equations is constructed out of the
energy-momentum tensor. In this article, we will investigate an idea such that a scale-invariant gravity
induces the spontaneous symmetry breakdown of gauge symmetry without assuming the existence of
the Higgs potential.

The SM based on $SU(3) \times SU(2) \times U(1)$ gauge group, together with classical general relativity, 
describes with amazing parsimony (only $19$ parameters) our world over scales that have been explored 
by experiments: from the Hubble radius of $10^{30} cm$ all the way down to scales of the order of
$10^{-16} cm$. In other words, with the help of cosmological initial conditions when the universe was
much smaller, the SM is believed to encode the information needed to deduce all the physical phenomena
observed so far. There are, however, some obvious chinks in the armor of the SM. In particular, the origin
of different scales in nature cannot be answered at all by the SM. One should recall that there is only
one fundamental constant with the dimension of mass: Gravity comes with its own mass scale 
$M_p = 2.4 \times 10^{18} GeV$. All units of mass should be scaled to this fundamental scale.
It is a source of great intellectual worry that the SM appears to be consistent at a scale which
is so different from the Planck mass scale. Naive expectations are that all physical phenomena
should occur at their natural scale which is of course the Planck scale.   

Coleman-Mandula theorem \cite{Mandula} allows the Poincare group to be generalized to two global groups,
one is the super-Poincare group and the other is the conformal group. It is remarkable to notice 
that these two groups might yield resolution of the gauge hierarchy problem by a completely 
different idea, and they also yield a natural generalization of local gauge group, the former
gives rise to the local super-Poincare group leading to supergravity whereas the latter 
does the local conformal group leading to conformal gravity.

According to recent results by the LHC \cite{ATLAS, CMS}, supersymmetry on the basis of the super-Poincare group
seems not to be taken by nature as resolution of the gauge hierarchy problem. Then, it is
natural to ask ourselves if  the conformal group, the other extension of the Poincare group, 
gives us resolution of  the gauge hierarchy problem.  Indeed, inspired by an interesting idea by
Bardeen \cite{Bardeen}, there has appeared to pursue the possibility of replacing the supersymmetry with
the conformal symmetry near the TeV scale in an attempt to solve the hierarchy problem \cite{Meissner, Iso}.
It is worth noting that the principle of conformal invariance is more rigid than the supersymmetry
in the sense that in many examples the conformal symmetry predicts the number of generations
as well as a rich structure for the Yukawa couplings among various families. This inter-family rigidity
is a welcome feature of the conformal approach to particle phenomenology \cite{Frampton}.

In the conformal approach, it is thought that the electro-weak scale and the QCD scale
as well as the masses of observed quarks and leptons are all so small compared to the
Planck scale that it is reasonable to believe that in some approximation they are exactly
massless. If so, then the quantum field theory which would be describing the massless fields 
should be a conformal theory as it has no mass scale. In this scenario, the fact that there are
no large mass corrections follows from the condition of conformal invariance. In other words,
the 'tHooft naturalness condition \cite{'tHooft} is satisfied in the conformal approach, namely in the
absence of masses there is an enhanced symmetry which is the conformal symmetry.
Of course, the breaking of conformal invariance should be soft in such a way that the idea of the
conformal symmetry is relevant for solving the hierarchy problem.

In passing, in the present context, it seems to be of interest to consider the issue of renormalizability.
Usually, in quantum field theories, the condition of renormalizability is imposed on a
theory as if it were a basic principle to make the perturbation method to be meaningful, but
its real meaning is unclear since there might exist a theory for which only the non-perturbative 
approach could be applied without relying on the perturbation method at all. To put differently,
the concept of renormalizability means that even if one is unfamiliar with true physics beyond
some higher energy scale, one can construct an effective theory by confining its ignorance 
to some parameters such as coupling constants and masses below the energy scale.
Thus, from this point of view, it is unclear to require the renormalizability to theories holding
at the highest energy scale, the Planck scale, such as quantum gravity and superstring theory.
On the other hand, given a scale invariance in a theory, all the coupling constants must
be dimensionless and operators in an action are marginal ones whose coefficient is independent
of a certain scale, which ensures that the theory is manifestly renormalizable. In this world,
all masses of particles must be then generated by spontaneous symmetry breakdown.

In previous works \cite{Oda1, Oda2}, we have shown that without resort to the Coleman-Weinberg mechanism
\cite{Coleman}, by coupling the non-minimal term of gravity, the U(1) B-L gauge symmetry in the model \cite{Iso}
is spontaneously broken in the process of spontaneous symmetry breakdown of global or local scale symmetry 
at the tree level and as a result the U(1) B-L gauge field becomes massive via the Higgs mechanism. One of advantages 
in this mechanism is that we do not have to introduce the Higgs potential in a theory.
  
Then, we have the following questions of this mechanism of symmetry breaking of gauge symmetry:
\begin{enumerate} 
\item Is it possible to generalize to the non-abelian gauge groups?   
\item Is it possible to generalize many scalar fields?
\item What becomes of applying it to the standard model and what its radiative corrections are?
\end{enumerate} 

In this article, we would like to answer these questions in order.
The structure of this article is the following: In Section 2, we present the simplest model which accomodates
global scale symmetry and the abelian gauge symmetry, and explain our main idea. 
In Section 3, we generalize this simple model to a model with the non-abelian gauge symmetry. 
In Section 4, we extend our idea to a model of a scalar field with many of components.
Moreover, we apply our finding to a scale-invariant extension of the standard model and calculate
radiative corrections in Section 5.  We conclude in Section 6. Two appendices are given, one
of which is to explain a derivation of the dilatation current and the other is to put useful
formulae for the calcualtion of radiative corrections.

\section{Review of a globally scale-invariant Abelian model}

We start with a brief review of the simplest model showing a gravitational Higgs phenomenon 
which was previously discovered in case of a global scale invariance and the abelian gauge group \cite{Oda1}.

With a background curved metric $g_{\mu\nu}$, a complex (singlet) scalar field $\Phi$ and
the $U(1)$ gauge field $A_\mu$, the Lagrangian takes the form\footnote{We follow 
notation and conventions by Misner et al.'s textbook \cite{MTW}, for instance, the flat Minkowski metric
$\eta_{\mu\nu} = diag(-, +, +, +)$, the Riemann curvature tensor $R^\mu \ _{\nu\alpha\beta} = 
\partial_\alpha \Gamma^\mu_{\nu\beta} - \partial_\beta \Gamma^\mu_{\nu\alpha} + \Gamma^\mu_{\sigma\alpha} 
\Gamma^\sigma_{\nu\beta} - \Gamma^\mu_{\sigma\beta} \Gamma^\sigma_{\nu\alpha}$, 
and the Ricci tensor $R_{\mu\nu} = R^\alpha \ _{\mu\alpha\nu}$.
The reduced Planck mass is defined as $M_p = \sqrt{\frac{c \hbar}{8 \pi G}} = 2.4 \times 10^{18} GeV$.
Through this article, we adopt the reduced Planck units where we set $c = \hbar = M_p = 1$ though we sometimes recover
the Planck mass $M_p$ for the clarification of explanation. In this units, all quantities become dimensionless. 
Finally, note that in the reduced Planck units, the Einstein-Hilbert Lagrangian density takes the form
${\cal L}_{EH} = \frac{1}{2} \sqrt{-g} R$.}:
\begin{eqnarray}
{\cal L} = \sqrt{-g} \left[ \xi \Phi^\dagger \Phi R - g^{\mu\nu} (D_\mu \Phi)^\dagger (D_\nu \Phi)
- \frac{1}{4} g^{\mu\nu} g^{\rho\sigma} F_{\mu\rho} F_{\nu\sigma} \right],
\label{Lagr 1}
\end{eqnarray}
where $\xi$ is a certain positive and dimensionless constant. The covariant derivative and field strength 
are respectively defined as 
\begin{eqnarray}
D_\mu \Phi = (\partial_\mu - i e A_\mu) \Phi,   \quad
(D_\mu \Phi)^\dagger = (\partial_\mu + i e A_\mu) \Phi^\dagger,   \quad
F_{\mu\nu} = \partial_\mu A_\nu - \partial_\nu A_\mu,
\label{Def 1}
\end{eqnarray}
with $e$ being a $U(1)$ real coupling constant.

Let us note that the Lagrangian (\ref{Lagr 1}) is invariant under a global scale transformation. In fact, 
with a constant parameter $\Omega = e^\Lambda \approx 1 + \Lambda \ (|\Lambda| \ll 1)$
the scale transformation is defined as \cite{Fujii1}\footnote{In this article, we use the terminology 
such that scale or conformal transformation means global
transformation whereas its local version is called local scale transformation or local conformal transformation.} 
\begin{eqnarray}
g_{\mu\nu} &\rightarrow& \tilde g_{\mu\nu} = \Omega^2 g_{\mu\nu},  \quad
g^{\mu\nu} \rightarrow \tilde g^{\mu\nu} = \Omega^{-2} g^{\mu\nu}, \quad
 \nonumber\\
\Phi &\rightarrow& \tilde \Phi = \Omega^{-1} \Phi, \quad
A_\mu \rightarrow \tilde A_\mu = A_\mu.
\label{Scale transf}
\end{eqnarray}
Then, using the formulae $\sqrt{-g} = \Omega^{-4} \sqrt{- \tilde g}, R = \Omega^2 \tilde R$, it is straightforward to
show that ${\cal L}$ is invariant under the scale transformation (\ref{Scale transf}).
Following the Noether procedure $\Lambda J^\mu = \sum \frac{\partial {\cal L}}{\partial \partial_\mu \phi} \delta \phi$ where 
$\phi = \{g_{\mu\nu}, \Phi, \Phi^\dagger \}$, as shown in the Appendix A, the current for the scale transformation, what
we call the dilatation current,  takes 
the form\footnote{The case $\xi = - \frac{1}{6}$ corresponds to conformal gravity, for which there is no dilatation current.}
\begin{eqnarray}
J^\mu =  ( 6 \xi + 1 ) \sqrt{-g} g^{\mu\nu} \partial_\nu \left(\Phi^\dagger \Phi \right).
\label{Current}
\end{eqnarray}

To prove that this current is conserved on-shell, it is necessary  to derive a set of equations of motion from
the Lagrangian (\ref{Lagr 1}). 
The variation of (\ref{Lagr 1}) with respect to the metric tensor produces Einstein's equations   
\begin{eqnarray}
2 \xi \Phi^\dagger \Phi G_{\mu\nu} = T^{(A)}_{\mu\nu} + T^{(\Phi)}_{\mu\nu} - 2 \xi ( g_{\mu\nu} \Box - \nabla_\mu \nabla_\nu )
(\Phi^\dagger \Phi),      
\label{Einstein eq}
\end{eqnarray}
where d'Alembert operator $\Box$ is as usual defined as $\Box (\Phi^\dagger \Phi) = \frac{1}{\sqrt{-g}} \partial_\mu
(\sqrt{-g} g^{\mu\nu} \partial_\nu (\Phi^\dagger \Phi)) = g^{\mu\nu} \nabla_\mu \nabla_\nu (\Phi^\dagger \Phi)$
and the Einstein tensor is $G_{\mu\nu} = R_{\mu\nu} - \frac{1}{2} g_{\mu\nu} R$.
Here the energy-momentum tensors $T^{(A)}_{\mu\nu}$ for the gauge field and $T^{(\Phi)}_{\mu\nu}$ 
for the scalar field are defined as, respectively
\begin{eqnarray}
T^{(A)}_{\mu\nu} &=& - \frac{2}{\sqrt{-g}} \frac{\delta}{\delta g^{\mu\nu}} 
[ - \frac{1}{4} \sqrt{-g} g^{\alpha\beta} g^{\rho\sigma} F_{\alpha\rho} F_{\beta\sigma} ]
\nonumber\\
&=&  g^{\rho\sigma} F_{\mu\rho} F_{\nu\sigma} - \frac{1}{4} g_{\mu\nu} F_{\rho\sigma}^2,   \nonumber\\
T^{(\Phi)}_{\mu\nu} &=& - \frac{2}{\sqrt{-g}} \frac{\delta}{\delta g^{\mu\nu}} [ - \sqrt{-g} g^{\rho\sigma} (D_\rho \Phi)^\dagger 
(D_\sigma \Phi) ]  \nonumber\\
&=&  2 (D_{(\mu} \Phi)^\dagger (D_{\nu)} \Phi) - g_{\mu\nu} (D_{\rho} \Phi)^\dagger (D^{\rho} \Phi),
\label{Energy-momentum}
\end{eqnarray}
where we have used the notation of symmetrization $A_{(\mu} B_{\mu)} = \frac{1}{2} (A_\mu B_\nu + A_\nu B_\mu)$. 

Next, the equation of motion for $\Phi^\dagger$ is of form
\begin{eqnarray}
\xi \Phi R  + \frac{1}{\sqrt{-g}} D_\mu (\sqrt{-g} g^{\mu\nu} D_\nu \Phi)  = 0.
\label{Phi eq}
\end{eqnarray}
Finally, taking the variation with respect to the gauge fields $A^{(i)}_\mu$ produces "Maxwell" equations
\begin{eqnarray}
\nabla_\rho F^{\mu\rho} = - i e \left[ \Phi^\dagger (D^\mu \Phi) - \Phi (D^\mu \Phi)^\dagger \right].
\label{Maxwell eq}
\end{eqnarray}

Now we wish to prove that the current (\ref{Current}) for the scale transformation is indeed conserved on-shell
by using these equations of motion.
Before doing so, let us first take the divergence of the current, whose result is given by  
\begin{eqnarray}
\partial_\mu J^\mu = ( 6 \xi + 1 ) \sqrt{-g} \Box (\Phi^\dagger \Phi).
\label{Div-Current}
\end{eqnarray}
In order to show that the expression in the right-hand side of Eq. (\ref{Div-Current}) vanishes on-shell, 
let us take the trace of Einstein's equations (\ref{Einstein eq})
\begin{eqnarray}
\xi \Phi^\dagger \Phi R = 3 \xi \Box (\Phi^\dagger \Phi) + (D_\mu \Phi)^\dagger (D^\mu \Phi).
\label{Trace-Einstein eq}
\end{eqnarray}
Next, multiplying Eq. (\ref{Phi eq}) by $\Phi^\dagger$, and then eliminating the term involving
the scalar curvature, i.e., $\xi \Phi^\dagger \Phi R$, with the help of Eq. (\ref{Trace-Einstein eq}), we obtain
\begin{eqnarray}
3 \xi \Box (\Phi^\dagger \Phi) + (D_\mu \Phi)^\dagger (D^\mu \Phi)
+ \frac{1}{\sqrt{-g}} \Phi^\dagger D_\mu (\sqrt{-g} g^{\mu\nu} D_\nu \Phi) = 0.
\label{Combined eq}
\end{eqnarray}

At this stage, it is useful to introduce a generalized covariant derivative defined
as ${\cal D}_\mu = D_\mu + \Gamma_\mu$ where $\Gamma_\mu$ is the usual affine connection. 
Using this derivative, Eq. (\ref{Combined eq}) can be rewritten as
\begin{eqnarray}
3 \xi {\cal D}_\mu {\cal D}^\mu (\Phi^\dagger \Phi) + ({\cal D}_\mu \Phi)^\dagger ({\cal D}^\mu \Phi)
+ \Phi^\dagger {\cal D}_\mu {\cal D}^\mu \Phi = 0.
\label{Re-Combined eq}
\end{eqnarray}
Then, adding its Hermitian conjugation to Eq. (\ref{Re-Combined eq}), we arrive at 
\begin{eqnarray}
(6 \xi + 1) {\cal D}_\mu {\cal D}^\mu (\Phi^\dagger \Phi) = 0.
\label{Re-Combined eq2}
\end{eqnarray}
The quantity $\Phi^\dagger \Phi$ is a scalar and neutral under the U(1) charge, 
we obtain 
\begin{eqnarray}
 (6 \xi + 1) \Box (\Phi^\dagger \Phi) = 0.
\label{Re-Combined eq3}
\end{eqnarray}
Using this equation, the right-hand side in Eq. (\ref{Div-Current}) is certainly vanishing, by which
we can prove that the current of the scale transformation is conserved on-shell
as promised.

Now we are willing to explain our finding about spontaneous symmetry breakdown of gauge symmetry
in our model where the coexistence of both scale invariance and gauge symmetry plays a pivotal role. 
Incidentally, it might be worthwhile to comment that in ordinary examples of spontaneous symmetry breakdown 
in the framework of quantum field theories, one is accustomed to dealing with a potential which has the shape of 
the Mexican hat type and therefore induces the symmetry breaking in a natural way, but
the same recipe cannot be applied to general relativity because of the lack of such a
potential.\footnote{In the case of massive gravity, a similar situation occurs in breaking
the general coordinate invariance spontaneously \cite{Oda}.}

Let us note that a very interesting recipe which induces spontaneous symmetry breakdown of $\it{scale}$
invariance via local scale transformation has been already known \cite{Fujii1}. 
This recipe can be explained as follows: Suppose that we started with 
a scale-invariant theory with only dimensionless coupling constants. But in the process
of local scale transformation, one cannot refrain from introducing the quantity with mass
dimension, which is the Planck mass $M_p$ in the present context, to match the dimensions 
of an equation and consequently scale invariance is spontaneously broken. 

Of course, the absence of a potential which induces symmetry breaking makes it impossible to
investigate a stability of the selected solution, but the very existence of the solution
including the Planck mass with mass dimension justifies the claim that this phenomenon is
nothing but a sort of spontaneous symmetry breakdown. This fact can be also understood 
by using a dilatation charge as seen shortly.

The first technique for obtaining spontaneous symmetry breakdown of both scale and gauge invariances is to 
find a suitable local scale transformation which transforms dilaton gravity in the Jordan frame to general relativity 
with matters in the Einstein frame. Of course, note that our starting Lagrangian is invariant under not the local scale transformation
but the global transformation, so the change of form of the Lagrangian after the local scale transformation is reasonable. 
Here it is useful to parametrize the complex scalar field $\Phi$ in terms of 
two real fields, $\Omega$ (or $\sigma$) and $\theta$ in polar form, defined as
\begin{eqnarray}
\Phi(x) = \frac{1}{\sqrt{2 \xi}} \Omega(x) e^{i \alpha \theta(x)} 
=  \frac{1}{\sqrt{2 \xi}} e^{\zeta \sigma(x) + i \alpha \theta(x)},
\label{Parametrization}
\end{eqnarray}
where $\Omega(x) = e^{\zeta \sigma(x)}$ is a local parameter field and the constants $\zeta, \alpha$ will be 
determined later.

Let us then consider the following local scale transformation:
\begin{eqnarray}
g_{\mu\nu} \rightarrow \tilde g_{\mu\nu} = \Omega^2(x) g_{\mu\nu},  \quad
g^{\mu\nu} \rightarrow \tilde g^{\mu\nu} = \Omega^{-2}(x) g^{\mu\nu}, \quad
A_\mu \rightarrow \tilde A_\mu = A_\mu.
\label{L-scale transf}
\end{eqnarray}
Note that apart from the local property of $\Omega(x)$, this local scale transformation is
different from the scale transformation (\ref{Scale transf}) in that the complex scalar
field $\Phi$ is not transformed at all. Under the local scale transformation (\ref{L-scale transf}),
the scalar curvature is transformed as
\begin{eqnarray}
R = \Omega^2 ( \tilde R + 6 \tilde \Box f - 6 \tilde g^{\mu\nu} \partial_\mu f \partial_\nu f ),
\label{Curvature}
\end{eqnarray}
where we have defined as $f = \log \Omega = \zeta \sigma$ and $\tilde \Box f = \frac{1}{\sqrt{- \tilde g}} 
\partial_\mu (\sqrt{- \tilde g} \tilde g^{\mu\nu} \partial_\nu f) = \tilde g^{\mu\nu} 
\tilde \nabla_\mu \tilde \nabla_\nu f$.

With the critical choice
\begin{eqnarray}
\xi \Phi^\dagger \Phi = \frac{1}{2} \Omega^2 = \frac{1}{2} e^{2 \zeta \sigma},
\label{Choice}
\end{eqnarray}
the non-minimal term in (\ref{Lagr 1}) reads the Einstein-Hilbert term (plus part of the kinetic term of 
the scalar field $\sigma$) up to a surface term as follows:
\begin{eqnarray}
\sqrt{-g} \xi \Phi^\dagger \Phi R &=& \Omega^{-4} \sqrt{- \tilde g} \frac{1}{2} \Omega^2
\Omega^2 ( \tilde R + 6 \tilde \Box f - 6 \tilde g^{\mu\nu} \partial_\mu f \partial_\nu f )  \nonumber\\
&=& \sqrt{- \tilde g} \left( \frac{1}{2} \tilde R - 3 \zeta^2 \tilde g^{\mu\nu} \partial_\mu \sigma \partial_\nu \sigma \right).
\label{1st term}
\end{eqnarray}
Then, the second term in (\ref{Lagr 1}) is cast to the form
\begin{eqnarray}
- \sqrt{-g} g^{\mu\nu} (D_\mu \Phi)^\dagger (D_\nu \Phi) 
= - \frac{1}{2 \xi} \sqrt{- \tilde g} \tilde g^{\mu\nu} \left( \zeta^2
\partial_\mu \sigma \partial_\nu \sigma + e^2 B_\mu B_\nu \right),
\label{2nd term}
\end{eqnarray}
where we have chosen $\alpha = e$ for convenience,
and defined a new massive gauge field $B_\mu$ as
\begin{eqnarray}
B_\mu = A_\mu + \partial_\mu \theta.
\label{B-field}
\end{eqnarray}
In terms of this new gauge field $B_\mu$, the Maxwell's Lagrangian in  (\ref{Lagr 1})
is described in the Einstein frame as
\begin{eqnarray}
- \frac{1}{4} \sqrt{-g} g^{\mu\nu} g^{\rho\sigma} F_{\mu\rho} F_{\nu\sigma} 
= - \frac{1}{4} \sqrt{- \tilde g} \tilde g^{\mu\nu} \tilde g^{\rho\sigma} \tilde F_{\mu\rho} \tilde F_{\nu\sigma},
\label{Maxwell Lagr}
\end{eqnarray}
where $\tilde F_{\mu\nu} \equiv \partial_\mu B_\nu - \partial_\nu B_\mu$.

It is worthwhile to stress again that in the process of local scale transformation we have had to introduce
the mass scale into a theory having no dimensional constants, thereby inducing the breaking of the scale
invariance. More concretely, to match the dimensions in the both sides of the equation,
the Planck mass $M_p$ must be introduced in the ciritical choice (\ref{Choice}) 
(recovering the Planck mass)
\begin{eqnarray}
\xi \Phi^\dagger \Phi = \frac{1}{2} \Omega^2 M_p^2 = \frac{1}{2} e^{2 \zeta \sigma} M_p^2.
\label{Choice2}
\end{eqnarray}
It is also remarkable to notice that in the process of spontaneous symmetry breakdown of the scale invariance, 
the Nambu-Goldstone boson $\theta$ is absorbed into the $U(1)$ gauge field $A_\mu$ as a longitudinal mode 
and as a result $B_\mu$ acquires a mass, which is nothing but the Higgs mechanism! In other words,
the $U(1)$ gauge symmetry is broken at the same time and on the same energy scale that the scale symmetry is spontaneously broken.
The size of the mass $M_B$ of $B_\mu$ can be read off from (\ref{2nd term}) as $M_B = \frac{e}{\sqrt{\xi}} M_p$
which is also equal to the energy scale on which the scale invariance is broken. 

Putting (\ref{1st term}), (\ref{2nd term}) and (\ref{Maxwell Lagr}) together, and defining $\zeta^{-2} = 6 + \frac{1}{\xi}$ 
(by which the kinetic term for the $\sigma$ field becomes a canonical form), the Lagrangian (\ref{Lagr 1})
is reduced to the form
\begin{eqnarray}
{\cal L} = \sqrt{- \tilde g} \left[ \frac{1}{2} M_p^2 \tilde R - \frac{1}{2} \tilde g^{\mu\nu} \partial_\mu \sigma \partial_\nu \sigma
- \frac{1}{4} \tilde F_{\mu\nu}^2 - \frac{e^2}{2 \xi} M_p^2 B_\mu B^\mu \right],
\label{Lagr 2}
\end{eqnarray}
where we have recovered the Planck mass $M_p$ for clarity.
Let us note that the first term coincides with the Einstein-Hilbert term in general relativity, the second term
implies that the dilaton $\sigma$ is massless at the classical level, and the last two terms means that
the gauge field becomes massive via the new Higgs mechanism. 

As an interesting application of our finding to phenomenology, we can propose two scenarios at the different energy scales.
One scenario, which was adopted in case of the classically scale-invariant B-L model \cite{Iso, Oda1, Oda2}, is
the spontaneous symmetry breakdown at the TeV scale where $\frac{e}{\sqrt{\xi}} \approx 10^{-15}$, so the gravity
is in the strong coupling phase. The other scenario is to trigger the spontaneous symmetry breakdown of both
scale and gauge symmetry at the Planck scale, for which we take $\frac{e}{\sqrt{\xi}} \approx 1$ and the gravity
is in the weak coupling phase.

Finally, let us comment on the physical meaning of the dilaton $\sigma$. The dilaton is a massless particle 
and interact with the other fields only through the covariant derivative $\tilde D_\mu = D_\mu + \zeta (\partial_\mu \sigma)$,
but owing to its nature of the derivative coupling, at the low energy this coupling is 
so small that it is difficult to detect the dilaton experimentally. 

To understand the physical meaning of the dilaton more clearly, it is useful to evaluate the dilatation current 
$J^\mu$ in (\ref{Current}) in the Einstein frame. The result reads
\begin{eqnarray}
J^\mu = \frac{1}{\zeta} \sqrt{- \tilde g} \tilde g^{\mu\nu} \partial_\nu \sigma.
\label{Current2}
\end{eqnarray}
This is exactly the expected form of the current seen in the case of the conventional spontaneous 
symmetry breakdown, with $\frac{1}{\zeta}$ playing the role of the vacuum value of the
order parameter, and the dilaton $\sigma$ doing of the Nambu-Goldstone boson 
associated with the spontaneous symmetry breakdown of the scale invariance.
This result can be also reached by constructing the corresponding charge which is 
defined as $Q_D = \int d^3 x J^0$. Note that this charge does not annihilate the vacuum 
because of the linear form in $\sigma$
\begin{eqnarray}
Q_D | 0 > \neq 0.
\label{Vacuum}
\end{eqnarray}
Of course, it is also possible to show $\partial_\mu J^\mu = 0$  in terms of equations of motion in the Einstein frame
as proved in the Jordan frame before.  It therefore turns out that the dilaton $\sigma$ is indeed the Nambu-Goldstone
boson associated with spontaneous symmetry breakdown of the scale invariance. We will see later that although the dilaton
is massless at the classical level, the trace anomaly makes the dilaton be massive at the quantum level.

\section{The generalization to non-Abelian groups}

In this section, we wish to extend the present formalism to arbitrary non-Abelian gauge groups. For clarity, 
we shall consider only the $SU(2)$ gauge group with a complex $SU(2)$-doublet of scalar field
$\Phi^T = (\Phi_1, \Phi_2)$ since the generalization to a general non-Abelian gauge group is
straightforward.

Let us start with the $SU(2)$ generalization of the Lagrangian (\ref{Lagr 1})
\begin{eqnarray}
{\cal L} = \sqrt{-g} \left[ \xi \Phi^\dagger \Phi R - g^{\mu\nu} (D_\mu \Phi)^\dagger (D_\nu \Phi)
- \frac{1}{4} g^{\mu\nu} g^{\rho\sigma} F_{\mu\rho}^a F_{\nu\sigma}^a \right],
\label{NA-Lagr 1}
\end{eqnarray}
where $a$ is an $SU(2)$ index running over $1, 2, 3$, and the covariant derivative and field strength 
are respectively defined as 
\begin{eqnarray}
D_\mu \Phi &=& (\partial_\mu - i g \tau^a A_\mu^a) \Phi,   \quad
(D_\mu \Phi)^\dagger = (\partial_\mu + i g \tau^a A_\mu^a) \Phi^\dagger,   \quad \nonumber\\
F_{\mu\nu}^a &=& \partial_\mu A_\nu^a - \partial_\nu A_\mu^a + g \varepsilon^{abc} A_\mu^b A_\nu^c.
\label{NA-Def 1}
\end{eqnarray}
Here $g$ is an $SU(2)$ coupling constant (Do not confuse with the determinant of the metric tensor since
we use the same letter of the alphabet). Furthermore, the matrices $\tau^a$ are defined as half of the Pauli ones, i.e.,
$\tau^a = \frac{1}{2} \sigma^a$, so the following relations are satisfied:
\begin{eqnarray}
\{ \tau^a, \tau^b \} = \frac{1}{2} \delta^{ab},  \quad
[ \tau^a, \tau^b ] = i \varepsilon^{abc} \tau^c.
\label{tau-matrix}
\end{eqnarray}
In order to see the Higgs mechanism discussed in the previous section explicitly, it is
convenient to go to the unitary gauge. To do that, we first parametrize the scalar doublet as
\begin{eqnarray}
{\Phi(x)} = U^{-1}(x) \frac{1}{\sqrt{2 \xi}} \ e^{\zeta \sigma(x)}  \left(
    \begin{array}{c}
      0 \\
      1 \\
    \end{array}
  \right),
\label{U-gauge}
\end{eqnarray}
where a unitary matrix $U(x)$ is defined as $U(x) = e^{ -i \alpha \tau^a \theta^a(x)}$ with $\alpha$ being
a real number. Then, we will define new fields in the unitary gauge by
\begin{eqnarray}
\Phi^u(x) &=& U(x) \Phi(x) = \frac{1}{\sqrt{2 \xi}} \ e^{\zeta \sigma(x)}  \left(
    \begin{array}{c}
      0 \\
      1 \\
    \end{array}
  \right)
= \frac{1}{\sqrt{2 \xi}} \ \Omega(x)  \left(
    \begin{array}{c}
      0 \\
      1 \\
    \end{array}
  \right),    \nonumber\\
\tau^a B_\mu^a &=& U(x) \tau^a A_\mu^a U^{-1}(x) - \frac{i}{g} \partial_\mu U(x) U^{-1}(x).
\label{NA-new fields}
\end{eqnarray}
Using these new fields, after an easy calculation, we find the following relations
\begin{eqnarray}
D_\mu \Phi = U^{-1}(x) D_\mu \Phi^u,   \quad
F_{\mu\nu}^a F^{a \mu\nu} = F_{\mu\nu}^a(B) F^{a \mu\nu}(B),
\label{NA-Rel 1}
\end{eqnarray}
where $D_\mu \Phi^u$ and $F_{\mu\nu}^a(B)$ are respectively defined as
\begin{eqnarray}
D_\mu \Phi^u = (\partial_\mu - i g \tau^a B_\mu^a) \Phi^u,   \quad
F_{\mu\nu}^a(B) = \partial_\mu B_\nu^a - \partial_\nu B_\mu^a + g \varepsilon^{abc} B_\mu^b B_\nu^c.
\label{NA-Def 2}
\end{eqnarray}

To reach the desired Lagrangian, we can follow a perfectly similar path of argument to the case
of the Abelian gauge group in the previous section. In other words, we will take a critical choice
\begin{eqnarray}
\xi \Phi^{u \dagger} \Phi^u = \frac{1}{2} \Omega^2 M_p^2 = \frac{1}{2} e^{2 \zeta \sigma} M_p^2,
\label{NA-Choice}
\end{eqnarray}
and make use of the local scale transformation by the local parameter $\Omega(x)$ to
move from the Jordan frame to the Einstein frame. After performing this procedure, the
final Lagrangian reads
\begin{eqnarray}
{\cal L} = \sqrt{- \tilde g} \left[ \frac{1}{2} M_p^2 \tilde R - \frac{1}{2} \tilde g^{\mu\nu} \partial_\mu \sigma \partial_\nu \sigma
- \frac{1}{4} (\tilde F_{\mu\nu}^a)^2 - \frac{g^2}{8 \xi} M_p^2 B_\mu^a B^{a \mu} \right],
\label{NA-Lagr 2}
\end{eqnarray}
where $\tilde F_{\mu\nu}^a \equiv F_{\mu\nu}^a(B)$. The mass of the massive gauge field $B_{\mu\nu}^a$
is easily read off to be $M_B = \frac{g}{2 \sqrt{\xi}} M_p$. As in the Abelian group, we can see that
the massless dilaton $\sigma$ is the Nambu-Goldstone boson of spontaneous symmetry breakdown of 
scale symmetry by making the conserved dilatation current and its charge.

\section{The generalization to scalar field with many components}

Let us recall that the quantum field theory of a scalar field with many components goes in
much the same way as that of a single component except that new interesting internal symmetry
arises. This general fact is also valid even in the present formalism if we take a common "radial"
field in all the components. When we consider a general "radial" field, we must face an annoying issue of getting
the canonical kinetic term for the dilaton. The procedure of obtaining the canonical kinetic term
is just a problem of matrix diagonalization and is not in principle a problem, but the general treatment makes
our formalism very complicated. We will therefore focus on the case of the common "radial" field in
this section. In the next section, we will meet the same situation since we consider two scalar fields coupling
to a curvature scalar in the non-minimal manner, but a reasonable approximation can serve to avoid this 
annoying issue.

The starting Lagrangian is just a generalization of (\ref{Lagr 1}) to $n$ complex scalar fields,
or equivalently a complex scalar field with $n$ independent components $\Phi_i ( i = 1, 2, \cdots, n )$
\begin{eqnarray}
{\cal L} = \sum_{i=1}^{n} \sqrt{-g} \left[ \xi_i \Phi_i^\dagger \Phi_i R - g^{\mu\nu} (D_\mu \Phi_i)^\dagger (D_\nu \Phi_i)
- \frac{1}{4} g^{\mu\nu} g^{\rho\sigma} F_{\mu\rho}^{(i)} F_{\nu\sigma}^{(i)} \right],
\label{M-Lagr 1}
\end{eqnarray}
where $\xi_i$ are positive and dimensionless constants. The covariant derivative and field strength 
are respectively defined as 
\begin{eqnarray}
D_\mu \Phi_i = (\partial_\mu - i e_i A_\mu^{(i)}) \Phi_i,   \quad
(D_\mu \Phi_i)^\dagger = (\partial_\mu + i e_i A_\mu^{(i)}) \Phi_i^\dagger,   \quad
F_{\mu\nu}^{(i)} = \partial_\mu A_\nu^{(i)} - \partial_\nu A_\mu^{(i)}.
\label{M-Def 1}
\end{eqnarray}

Since the Lagrangian (\ref{M-Lagr 1}) includes only dimensionless coupling constants, it is manifestly
invariant under a global scale transformation. Following the Noether theorem, the current for the scale 
transformation reads
\begin{eqnarray}
J^\mu =  \sum_{i=1}^{n} ( 6 \xi_i + 1 ) \sqrt{-g} g^{\mu\nu} \partial_\nu \left(\Phi_i^\dagger \Phi_i \right).
\label{M-Current}
\end{eqnarray}
In a similar way to the cases of both Abelian and non-Abelian gauge groups, we can show that this current is conserved 
on-shell.

As mentioned in the above, the point is to take a common "radial" (real) field $\Omega(x)$ such that
\begin{eqnarray}
\Phi_i (x) = \frac{1}{\sqrt{2 n \xi_i}} \Omega(x) e^{i \alpha_i \theta_i (x)} 
=  \frac{1}{\sqrt{2 n \xi_i}} e^{\zeta \sigma(x) + i \alpha_i \theta_i (x)}.
\label{M-Parametrization}
\end{eqnarray}
This is a great simplification in the sense that $n$ real component fields in $\Phi_i (x)$ is reduced to
a single one, but this restriction is needed to obtain the canonical kinetic term for
the dilaton in a rather simple way. 

Now we would like to show that the Lagrangian (\ref{M-Lagr 1}) has the property of spontaneous symmetry breakdown
of gauge symmetry when scale symmetry is spontaneously broken. To do that, we proceed similar steps 
to the case of the single scalar field with the Abelian gauge group in Section 2.
With the critical choice
\begin{eqnarray}
\xi_i \Phi_i^\dagger \Phi_i = \frac{1}{2 n} \Omega^2 = \frac{1}{2 n} e^{2 \zeta \sigma},
\label{M-Choice}
\end{eqnarray}
the non-minimal term in (\ref{M-Lagr 1}) yields the Einstein-Hilbert term and part of the kinetic term of the 
scalar field $\sigma$) up to a surface term
\begin{eqnarray}
\sum_{i=1}^{n} \sqrt{-g} \xi_i \Phi_i^\dagger \Phi_i R
= \sqrt{- \tilde g} \left( \frac{1}{2} \tilde R - 3 \zeta^2 \tilde g^{\mu\nu} \partial_\mu \sigma \partial_\nu \sigma \right).
\label{M-1st term}
\end{eqnarray}
Moreover, the second term in (\ref{M-Lagr 1}) is reduced to
\begin{eqnarray}
- \sum_{i=1}^{n} \sqrt{-g} g^{\mu\nu} (D_\mu \Phi_i)^\dagger (D_\nu \Phi_i) 
= - \sum_{i=1}^{n} \frac{1}{2 n \xi_i} \sqrt{- \tilde g} \tilde g^{\mu\nu} \left( \zeta^2
\partial_\mu \sigma \partial_\nu \sigma + e_i^2 B_\mu^{(i)} B_\nu^{(i)} \right),
\label{M-2nd term}
\end{eqnarray}
where we have selected $\alpha_i = e_i$ and defined new massive gauge fields $B_\mu^{(i)}$ by
\begin{eqnarray}
B_\mu^{(i)} = A_\mu^{(i)} + \partial_\mu \theta_i.
\label{M-B-field}
\end{eqnarray}
In terms of the new gauge fields $B_\mu^{(i)}$, the Maxwell's Lagrangian in  (\ref{M-Lagr 1})
is cast to the form 
\begin{eqnarray}
- \frac{1}{4} \sum_{i=1}^{n} \sqrt{-g} g^{\mu\nu} g^{\rho\sigma} F_{\mu\rho}^{(i)} F_{\nu\sigma}^{(i)} 
= - \frac{1}{4} \sum_{i=1}^{n} \sqrt{- \tilde g} \tilde g^{\mu\nu} \tilde g^{\rho\sigma} \tilde F_{\mu\rho}^{(i)} 
\tilde F_{\nu\sigma}^{(i)},
\label{M-Maxwell Lagr}
\end{eqnarray}
where $\tilde F_{\mu\nu}^{(i)} \equiv \partial_\mu B_\nu^{(i)} - \partial_\nu B_\mu^{(i)}$.

To summarize, the Lagrangian (\ref{M-Lagr 1})  is given by
\begin{eqnarray}
{\cal L} = \sqrt{- \tilde g} \left\{ \frac{1}{2} M_p^2 \tilde R - \frac{1}{2} \tilde g^{\mu\nu} \partial_\mu \sigma \partial_\nu \sigma
+ \sum_{i=1}^{n} \left[ - \frac{1}{4} (\tilde F_{\mu\nu}^{(i)})^2 - \frac{e_i^2}{2 n \xi_i} M_p^2 B_\mu^{(i)} B^{(i) \mu} \right] \right\},
\label{M-Lagr 2}
\end{eqnarray}
where the definition $\zeta^{-2} = 6 + \frac{1}{n} \sum_{i=1}^{n} \frac{1}{\xi_i}$ is used. It is then obvious that this
system also exhibits the new Higgs mechanism triggered by spontaneous symmetry breakdown of scale
symmetry.

\section{Radiative corrections and dilaton mass}

In this section, as an example, we wish to apply our idea discussed so far to
the standard model and evaluate the quantum effects. Since the standard model
is known to not be classically scale-invariant because of the presence of the
(negative) mass term of the Higgs field, we must replace the mass term with
a new scalar field. It is then natural to identify this new scalar with the $\Phi$ field
which couples to a scalar curvature in a non-minimal manner.

Let us first recall that in our previous work \cite{Oda1} we have already considered 
one-loop effects of a classically scale invariant B-L model \cite{Iso}. 
However, our finding of gravitational spontaneous
symmetry breakdown of gauge symmetry as a result of spontaneous symmetry breakdown of
scale symmetry is very universal in the sense that our ideas can be generalized not only to
local scale symmetry as clarified in \cite{Oda2} but also to non-Abelian gauge groups and 
even a scalar field with many of components as discussed in this article.
In fact, it is obvious that our ideas can be applied to any model which is scale-invariant
and involves the non-minimal coupling terms between the curvature scalar and charged
scalars associated with local gauge symmetries.

In our calculation, we are not ambitious enough to quantize the metric tensor field and
take a fixed Minkowski background $g_{\mu\nu} = \eta_{\mu\nu}$. Moreover, we restrict
ourselves to the calculation of radiative corrections between dilaton and matter
fields in the weak-field approximation. 

One of the motivations behind this study is to calculate the size of the mass of dilaton. As shown
above, the dilaton is exactly massless at the classical level owing to scale symmetry, but
it is well-known that radiative corrections violate the scale invariance thereby leading to
the trace anomaly. Consequently, the dilaton becomes massive in the quantum regime.
Since the dilaton is a scalar field like the Higgs particle, one might expect that there
could be quadratic divergence for the self-energy diagram. 
On the other hand, since the dilaton is the Nambu-Goldstone boson resulting from the scale
symmetry, the dilaton mass would be much lower in the such a way that the pion
masses are very smaller as the pions can be understood as the Nambu-Goldstone boson
coming from $SU(2)_L \times SU(2)_R \rightarrow SU(2)_V$ flavor symmetry breaking.
As long as we know, nobody has calculated the dilaton mass in a reliable manner, so
we wish to calculate the dilaton mass within the framework of the present formalism
and determine which scenario, quadratic divergence and huge radiative corrections like the Higgs particle 
or very lower mass like the pions, is realized.  Remarkably enough, it will be shown that although
the dilaton mass is quadratic divergent, the cutoff scale, which is the Planck mass
in the formalism at hand, is exactly cancelled by the induced coupling constant, by which
the dilaton mass is kept to be around the GeV scale. 

Whenever we evaluate anomalies, the key point is to adopt a suitable regularization method
respecting classical symmetries existing in the action as much as possible. In this article,
as a regularization method, we make use of the method of continuous space-time dimensions,
for which we rewrite previous results in arbitrary $D$ dimensions \cite{Fujii2}.
Like the dimensional regularization, the divergences will appear as poles $\frac{1}{D-4}$,
which are cancelled by the factor $D-4$ that multiplies the dilaton coupling, thereby 
producing a finite result leading to an effective interaction term.

\subsection{Basic formalism}

Our starting Lagrangian, which is a scale-invariant extension of the standard model
coupled to the non-minimal terms, is of form
\begin{eqnarray}
{\cal L} &=& \sqrt{-g} \Big[ ( \xi_1 \Phi^\dagger \Phi + \xi_2 H^\dagger H ) R 
- g^{\mu\nu} (D_\mu \Phi)^\dagger (D_\nu \Phi) - g^{\mu\nu} (D_\mu H)^\dagger (D_\nu H)
\nonumber\\
&-& \frac{1}{4} g^{\mu\nu} g^{\rho\sigma} ( F_{\mu\rho}^{(1)} F_{\nu\sigma}^{(1)}
+ F_{\mu\rho}^{(2)} F_{\nu\sigma}^{(2)} + F_{\mu\rho}^a F_{\nu\sigma}^a )  
- V (H, \Phi) + L_m \Big],
\label{Q-Lagr 1}
\end{eqnarray}
where $L_m$ denotes the remaining Lagrangian part of the standard-model sector such as
the Yukawa couplings and various definitions are given by the following expressions:
\begin{eqnarray}
D_\mu \Phi &=& (\partial_\mu - i e_1 A_\mu^{(1)}) \Phi,   \quad
D_\mu H = (\partial_\mu - i g \tau^a A_\mu^a - i \frac{e_2}{2}  A_\mu^{(2)}) H,   \nonumber\\
F_{\mu\nu}^{(i)} &=& \partial_\mu A_\nu^{(i)} - \partial_\nu A_\mu^{(i)},    \quad
F_{\mu\nu}^a = \partial_\mu A_\nu^a - \partial_\nu A_\mu^a + g \varepsilon^{abc} A_\mu^b A_\nu^c,
\nonumber\\
V (H, \Phi) &=& \lambda_\Phi (\Phi^\dagger \Phi)^2 + \lambda_{H \Phi} (H^\dagger H) (\Phi^\dagger \Phi)
+ \lambda_H (H^\dagger H)^2. 
\label{Qb 1}
\end{eqnarray}
with $e_i (i = 1, 2)$ being $U(1)$ coupling constants and $g$ being an $SU(2)$ coupling constant.
As in the Appendix A, in this model, we can also calculate the Noether current for scale transformation
\begin{eqnarray}
J^\mu =  \sqrt{-g} g^{\mu\nu} \partial_\nu [ ( 6 \xi_1 + 1 )  \Phi^\dagger \Phi
+ ( 6 \xi_2 + 1 )  H^\dagger H ].
\label{Q-Current}
\end{eqnarray}
It turns out that this dilatation current is conserved on-shell as well. 

For simplicity of presentation, we take the vanishing $SU(2)$ gauge field, $A_\mu^a = 0$ since this
assumption does not change the essential conclusion for our purpose. 

In general $D$ space-time dimensions, as a generalization of Eq.  (\ref{L-scale transf}),
the local scale transformation is defined as
\begin{eqnarray}
\hat g_{\mu\nu} &=& \Omega^2(x) g_{\mu\nu},  \quad \hat g^{\mu\nu} = \Omega^{-2}(x) g^{\mu\nu}, \quad
\hat \Phi = \Omega^{- \frac{D-2}{2}}(x) \Phi, \nonumber\\
\hat H &=& \Omega^{- \frac{D-2}{2}}(x) H,  \quad \hat A^{(i)}_\mu = \Omega^{- \frac{D-4}{2}}(x) A^{(i)}_\mu.
\label{Q-L-scale transf}
\end{eqnarray}
Under this local scale transformation (\ref{Q-L-scale transf}), with the definition $f = \log \Omega$, 
the scalar curvature is transformed as
\begin{eqnarray}
R = \Omega^2 \left[ \hat R + 2 (D-1) \hat \Box f - (D-1) (D-2) \hat g^{\mu\nu} \partial_\mu f \partial_\nu f \right],
\label{Q-Curvature}
\end{eqnarray}
for which we set $D=4$ in what follows since we do not quantize the metric tensor and therefore do not have poles 
from the curvature. 

In a physically more realistic situation, the scale symmetry must be broken spontaneously in the higher energy
region before spontaneous symmetry breaking of the electro-weak symmetry since all quantum
field theories must in principle contain the gravity from the beginning although contributions from
the gravity can be usually ignored when dealing with particle physics processes.

Therefore, let us first break the scale invariance by taking the following value for the charged 
scalar field $\Phi$:   
\begin{eqnarray}
\Phi = \frac{1}{\sqrt{2 \xi_1}} \Omega^{\frac{D-2}{2}} e^{i \alpha \theta}
= \frac{1}{\sqrt{2 \xi_1}} e^{\zeta \sigma + i \alpha \theta},
\label{Q-Choice}
\end{eqnarray}
where we have defined $\Omega(x) = e^{\frac{2}{D-2} \zeta \sigma}$ and 
$\zeta^{-2} \equiv 4 \frac{D-1}{D-2} + \frac{1}{\xi_1} = 6 + \frac{1}{\xi_1}$.
Then, the first term in (\ref{Q-Lagr 1}) takes the form
\begin{eqnarray}
\sqrt{-g} \xi_1 \Phi^\dagger \Phi R = 
\sqrt{- \hat g} \left( \frac{1}{2} \hat R - 3 \zeta^2 \hat g^{\mu\nu} \partial_\mu \sigma \partial_\nu \sigma \right).
\label{Q-1st term}
\end{eqnarray}
Similarly, the third term in (\ref{Q-Lagr 1}) is changed to the form
\begin{eqnarray}
- \sqrt{-g} g^{\mu\nu} (D_\mu \Phi)^\dagger (D_\nu \Phi) 
= - \frac{1}{2 \xi_1} \sqrt{- \hat g} \hat g^{\mu\nu} \left( \zeta^2
\partial_\mu \sigma \partial_\nu \sigma + \hat e_1^2 \hat B_\mu^{(1)} \hat B_\nu^{(1)} \right),
\label{Q-3rd term}
\end{eqnarray}
where we have defined a new coupling constant and massive gauge field as  
\begin{eqnarray}
\hat e_1 = \Omega^{\frac{D-4}{2}} e_1, \quad \hat B_\mu^{(1)} = \hat A_\mu^{(1)} - \partial_\mu \theta,
\label{Q-B-field}
\end{eqnarray}
and chosen $\alpha = \hat e_1$ for convenience.
Adding (\ref{Q-1st term}) and (\ref{Q-3rd term}) together yields the expression
\begin{eqnarray}
\sqrt{-g} \left[ \xi_1 \Phi^\dagger \Phi R - g^{\mu\nu} (D_\mu \Phi)^\dagger (D_\nu \Phi) \right] 
= \sqrt{- \hat g} \left( \frac{1}{2} \hat R - \frac{1}{2} \hat g^{\mu\nu} \partial_\mu \sigma \partial_\nu \sigma
- \frac{\hat e_1^2}{2 \xi_1} \hat g^{\mu\nu} \hat B_\mu^{(1)} \hat B_\nu^{(1)} \right).
\label{Q-1st+3rd term}
\end{eqnarray}

On the other hand, the Lagrangian of matter fields turns out to depend on the dilaton field $\sigma$
in a non-trivial manner in general $D$ space-time dimensions.  First, the non-minimal term for $H$ field
becomes
\begin{eqnarray}
\sqrt{-g} \xi_2 H^\dagger H R
= \sqrt{- \hat g} \ \xi_2 \hat H^\dagger \hat H 
\left( \hat R + 6 \zeta^2 \hat \Box \sigma 
- 6 \zeta^2 \hat g^{\mu\nu} \partial_\mu \sigma \partial_\nu \sigma \right).
\label{Q-H}
\end{eqnarray}
Second, the kinetic term for $H$ is cast to
\begin{eqnarray}
- \sqrt{-g} g^{\mu\nu} (D_\mu H)^\dagger (D_\nu H)
= - \sqrt{- \hat g} \hat g^{\mu\nu} (\hat D_\mu \hat H)^\dagger (\hat D_\nu \hat H),
\label{Q-H2}
\end{eqnarray}
where the new covariant derivative is defined as
\begin{eqnarray}
\hat D_\mu \hat H = ( \partial_\mu + \zeta \partial_\mu \sigma 
- \frac{i}{2} \hat e_2 \hat A_\mu^{(2)} ) \hat H,
\label{Q-DH}
\end{eqnarray}
with being $\hat e_2 = \Omega^{\frac{D-4}{2}} e_2$.
Third, the electro-magnetic terms are reduced to the form
\begin{eqnarray}
- \frac{1}{4} \sqrt{-g} \sum_{i=1}^2 g^{\mu\nu} g^{\rho\sigma} F^{(i)}_{\mu\rho} F^{(i)}_{\nu\sigma}
= - \frac{1}{4} \sqrt{- \hat g} \sum_{i=1}^2 \hat g^{\mu\nu} \hat g^{\rho\sigma} 
\hat F^{(i)}_{\mu\rho} \hat F^{(i)}_{\nu\sigma},
\label{Q-EM}
\end{eqnarray}
where the new field strengths are defined as\footnote{The presence of the Nambu-Goldstone mode
$\theta$ in $\hat F^{(1)}_{\mu\nu}$ merely shows that scale invariance of the theory under
consideration is violated in any space-time dimension except four dimensions.}
\begin{eqnarray}
\hat F^{(1)}_{\mu\nu} &=& \Omega^{2- \frac{D}{2}} F^{(1)}_{\mu\nu} 
= \partial_\mu \hat B^{(1)}_\nu + \frac{D-4}{2} \zeta \partial_\mu \sigma (\hat B^{(1)}_\nu
+ \partial_\nu \theta) - (\mu \leftrightarrow \nu), \nonumber\\
\hat F^{(2)}_{\mu\nu} &=& \Omega^{2- \frac{D}{2}} F^{(2)}_{\mu\nu} 
= \partial_\mu \hat A^{(2)}_\nu + \frac{D-4}{2} \zeta \partial_\mu \sigma \hat A^{(2)}_\nu
- (\mu \leftrightarrow \nu).
\label{Q-F}
\end{eqnarray}
Finally, the potential term can be rewritten as
\begin{eqnarray}
\sqrt{-g} V (H, \Phi)  &=& \sqrt{- \hat g} V(\hat H)  \nonumber\\
&=& \sqrt{- \hat g} e^{\frac{2(D-4)}{D-2} \zeta \sigma}  \left[ \frac{1}{4 \xi_1^2} \lambda_\Phi M_p^4 
+ \frac{1}{2 \xi_1} \lambda_{H \Phi} M_p^2 (\hat H^\dagger \hat H)
+ \lambda_H (\hat H^\dagger \hat H)^2 \right].
\label{Q-P}
\end{eqnarray}
To summarize, the starting Lagrangian is now of the form
\begin{eqnarray}
{\cal L} &=& \sqrt{- \hat g} \Big[  \frac{1}{2} \hat R 
- \frac{1}{2} \hat g^{\mu\nu} \partial_\mu \sigma \partial_\nu \sigma
- \frac{\hat e_1^2}{2 \xi_1} \hat g^{\mu\nu} \hat B_\mu^{(1)} \hat B_\nu^{(1)} 
\nonumber\\
&+& \xi_2 \hat H^\dagger \hat H \left( \hat R + 6 \zeta^2 \hat \Box \sigma 
- 6 \zeta^2 \hat g^{\mu\nu} \partial_\mu \sigma \partial_\nu \sigma \right)
- \hat g^{\mu\nu} (\hat D_\mu \hat H)^\dagger (\hat D_\nu \hat H)
\nonumber\\
&-& \frac{1}{4} \sum_{i=1}^2 \hat g^{\mu\nu} \hat g^{\rho\sigma} 
\hat F^{(i)}_{\mu\rho} \hat F^{(i)}_{\nu\sigma}
- V (\hat H) + L_m \Big].
\label{Q-Lagr 2}
\end{eqnarray}

Next, we are ready to deal with spontaneous symmetry breakdown of the electro-weak symmetry,
which is assumed to occur at the lower energy, GeV scale, than breaking of scale symmetry. 
To realize the spontaneous symmetry breakdown of the electro-weak symmetry, we assume the
conventional ansatz
\begin{eqnarray}
\lambda_{H \Phi} < 0, \quad \lambda_{H} > 0.
\label{Q-Ansatz}
\end{eqnarray}
With the parametrization $\hat H^T = (0, \frac{v + h}{\sqrt{2}}) e^{i \varphi}$, after 
the spontaneous symmetry breakdown of the electro-weak symmetry, the potential term
can be described as  
\begin{eqnarray}
V(\hat H) 
= e^{\frac{2(D-4)}{D-2} \zeta \sigma}  \left[ \frac{1}{2} m_h^2 h^2 + \sqrt{\frac{\lambda_H}{2}} m_h h^3
+ \frac{\lambda_H}{4} h^4 \right],
\label{Q-P2}
\end{eqnarray}
where a constant in the square bracket is discarded, and the vacuum expectation value $v$ and
the Higgs mass $m_h$ are respectively defined as
\begin{eqnarray}
v^2 = \frac{1}{2 \xi_1} \frac{|\lambda_{H \Phi}|}{\lambda_H} M_p^2, \quad
m_h = \sqrt{2 \lambda_H} v = \sqrt{\frac{|\lambda_{H \Phi}|}{\xi_1}} M_p,
\label{Q-P3}
\end{eqnarray}
where the Planck mass $M_p$ is explicitly written.
 
Now we wish to consider couplings between the dilaton field $\sigma$ and matter fields which
vanish at the classical level ($D=4$) but provide a finite contribution at the quantum level, interpreted as the
trace anomaly. In the weak field approximation, let us extract terms linear in the dilaton $\sigma$ in 
$V(\hat H)$ as
\begin{eqnarray}
e^{\frac{2(D-4)}{D-2} \zeta \sigma} \approx 1 + (D-4) \zeta \sigma.
\label{Q-EXP}
\end{eqnarray}
Then, the potential $V(\hat H)$ is devided into two parts  
\begin{eqnarray}
V(\hat H) = V^{(0)} (\hat H) + V^{(1)} (\hat H), 
\label{Q-P4}
\end{eqnarray}
where we have defined as
\begin{eqnarray}
V^{(0)} (\hat H) &=& \frac{1}{2} m_h^2 h^2 + \sqrt{\frac{\lambda_H}{2}} m_h h^3 
+ \frac{\lambda_H}{4} h^4,
\nonumber\\
V^{(1)} (\hat H) &=& (D-4) \zeta V^{(0)} (\hat H) \sigma.
\label{Q-P5}
\end{eqnarray}

Using the parametrization $\hat H^T = (0, \frac{v + h}{\sqrt{2}}) e^{i \varphi}$, the remaining
part including the field $H$ except the Higgs potential is also rewritten, and consequently
the whole Lagrangian (\ref{Q-Lagr 2}) takes a little longer expression
\begin{eqnarray}
{\cal L} &=& \sqrt{- \hat g} \ \Big\{  \frac{1}{2} M_p^2 \hat R 
- \frac{1}{2} \hat g^{\mu\nu} \partial_\mu \sigma \partial_\nu \sigma
- \frac{\hat e_1^2}{2 \xi_1} M_p^2 \hat g^{\mu\nu} \hat B_\mu^{(1)} \hat B_\nu^{(1)} 
\nonumber\\
&+& \frac{1}{2} \xi_2 v^2 \hat R + \xi_2 (v h + \frac{1}{2} h^2) ( \hat R - 6 \zeta^2 \frac{1}{M_p^2}
\hat g^{\mu\nu} \partial_\mu \sigma \partial_\nu \sigma )
\nonumber\\
&-& 3 \xi_2 \zeta^2 \frac{v^2}{M_p^2} \hat g^{\mu\nu} \partial_\mu \sigma \partial_\nu \sigma
-6 \xi_2 \zeta^2 \frac{v}{M_p} \hat g^{\mu\nu} \partial_\mu h \partial_\nu \sigma
+ 3 \xi_2 \zeta^2 \frac{1}{M_p} h^2 \hat \Box \sigma
\nonumber\\
&-& \hat g^{\mu\nu} \Big[ \frac{1}{2} \partial_\mu h \partial_\nu h 
+ \zeta \frac{1}{M_p} \partial_\mu h \partial_\nu \sigma ( v + h ) 
+ \frac{1}{2} \zeta^2 \frac{1}{M_p^2} \partial_\mu \sigma \partial_\nu \sigma ( v + h )^2 
\nonumber\\
&+& \frac{\hat e_2^2}{8} \hat A_\mu^{(2)} \hat A_\nu^{(2)} ( v + h )^2 \Big]
- \frac{1}{4} \sum_{i=1}^2 \hat g^{\mu\nu} \hat g^{\rho\sigma} \hat F^{(i)}_{\mu\rho} \hat F^{(i)}_{\nu\sigma}
- V (\hat H) + L_m \Big\},
\label{Q-Lagr 3}
\end{eqnarray}
where we have recovered the Planck mass scale $M_p$.
As mentioned in Section 4, given two non-minimal terms, we need to diagonalize the kinetic terms for the
dilaton $\sigma$ and the Higgs field $h$ to get the canonical form. However, in the present context,
the energy scale $v$ of the electro-weak symmetry breaking is much lower compared to that
of the scale symmetry one, so it is reasonable to take the approximation 
\begin{eqnarray}
\frac{v}{M_p} \ll 1, \quad \xi_2 v^2 \ll M_p^2.
\label{Q-Approx}
\end{eqnarray}
With this approximation, the Lagrangian is rather simplified to
\begin{eqnarray}
{\cal L} &=& \sqrt{- \hat g} \ \Big[  \frac{1}{2} \hat R 
- \frac{1}{2} \hat g^{\mu\nu} \partial_\mu \sigma \partial_\nu \sigma
- \frac{\hat e_1^2}{2 \xi_1} \hat g^{\mu\nu} \hat B_\mu^{(1)} \hat B_\nu^{(1)} 
\nonumber\\
&+& \xi_2 (v h + \frac{1}{2} h^2) \hat R 
- \frac{1}{2} \hat g^{\mu\nu} \partial_\mu h \partial_\nu h 
- \frac{\hat e_2^2}{8} \hat g^{\mu\nu} \hat A_\mu^{(2)} \hat A_\nu^{(2)} ( v + h )^2 
\nonumber\\
&-& \frac{1}{4} \sum_{i=1}^2 \hat g^{\mu\nu} \hat g^{\rho\sigma} \hat F^{(i)}_{\mu\rho} \hat F^{(i)}_{\nu\sigma}
- V (\hat H) + L_m \Big].
\label{Q-Lagr 4}
\end{eqnarray}
Based on this Lagrangian, we wish to calculate quantum effects, in particular, on the dilaton
coupling below. Since we are interested in the low energy region, the derivative coupling of the dilaton
appearing in $\hat F_{\mu\nu}^{(i)}$ and $\hat D_\mu \hat H$ will be ignored in the calculation.

\subsection{The coupling between dilaton and Higgs field}

In this subsection, we first switch off the $U(1)$ fields and calculate the coupling between the dilaton $\sigma$ 
and the Higgs particle $h$ and derive an effective Lagrangian at the one-loop level. The contribution from
the $U(1)$ fields will be discussed in the later subsection. We will see that the $\sigma h^n (2 \le n \le 4)$
$(n+1)$-point diagrams are non-vanishing whereas the $\sigma h^n (n \ge 5)$ diagrams are vanishing.
  
First, let us consider three-point (with two Higgs $h$ and one dilaton $\sigma$ as the external particles),
one-loop diagrams. 
Inspection of the vertices reveals that we have three types of one-loop divergent diagrams in which
the Higgs field is circulating in the loop and one dilaton field, whose momentum is assumed to be vanishing, couples. 
Note that the divergences stemming from the Higgs one-loop diagrams provide us with poles $\frac{1}{D-4}$,
which cancel the factor $D-4$ multiplying the dilaton coupling in $V^{(1)} (\hat H)$, thereby yielding
a finite contribution. 

One type of one-loop divergent diagram, which we call the diagram (A1), is a tadpole type and is
given by the Higgs loop to which the dilaton couples by the vertex $- (D-4) 3! \zeta \lambda_H$ in $V^{(1)} (\hat H)$. 
The corresponding amplitude ${\cal{T}}_{A1}$ is of form 
\begin{eqnarray}
{\cal{T}}_{A1} &=& - i (D-4) 3! \zeta \lambda_H  \int \frac{d^D k}{(2 \pi)^D} \frac{1}{k^2 + m_h^2}
\nonumber\\
&=& - i (D-4) 3! \zeta \lambda_H \frac{i \pi^2}{(2 \pi)^4} (m_h^2)^{\frac{D}{2} - 1} \Gamma(1 - \frac{D}{2})
\nonumber\\
&=& \frac{3}{4 \pi^2} \zeta  \lambda_H  m_h^2,
\label{T-A1-1}
\end{eqnarray}
where we have used  the familiar formula in the dimensional regularization which corresponds to a specific case 
of the general formula in Appendix B
\begin{eqnarray}
\int \frac{d^D k}{(2 \pi)^D} \frac{1}{k^2 + m_h^2}
= \frac{i \pi^2}{(2 \pi)^4} (m_h^2)^{\frac{D}{2} -1} \Gamma(1- \frac{D}{2}),
\label{Q-F1}
\end{eqnarray}
and the property of the gamma function $\Gamma(m+1) = m \Gamma(m)$.

The second type of one-loop divergent diagram, which we call the diagram (A2), is given by the Higgs loop 
to which the dilaton couples by the vertex $- (D-4) \zeta m_h^2$ in $V^{(1)} (\hat H)$ and
with the Higgs self-coupling vertex $- 3! \lambda_H$ in $V^{(0)} (\hat H)$. 
The amplitude ${\cal{T}}_{A2}$ is calculated as 
\begin{eqnarray}
{\cal{T}}_{A2} &=& i (D-4) 3! \zeta \lambda_H m_h^2 \int \frac{d^D k}{(2 \pi)^D} \frac{1}{(k^2 + m_h^2)^2}
\nonumber\\
&=& i (D-4) 3! \zeta \lambda_H m_h^2 \frac{i}{16 \pi^2} \Gamma(2 - \frac{D}{2})
\nonumber\\
&=& \frac{3}{4 \pi^2} \zeta  \lambda_H  m_h^2,
\label{T-A2-1}
\end{eqnarray}
where we have used  the equation
\begin{eqnarray}
\int \frac{d^D k}{(2 \pi)^D} \frac{1}{(k^2 + m_h^2)^2}
&=& - \frac{\partial}{\partial m_h^2} \int \frac{d^D k}{(2 \pi)^D} \frac{1}{k^2 + m_h^2}
\nonumber\\
&=& \frac{i \pi^2}{(2 \pi)^4} (m_h^2)^{\frac{D}{2} - 2} (1 - \frac{D}{2}) \Gamma(1- \frac{D}{2})
\nonumber\\
&=& \frac{i}{16\pi^2} \Gamma(2- \frac{D}{2}).
\label{Q-F2}
\end{eqnarray}

The final type of one-loop diagram, which we call the diagram (A3), is a little more involved and given by the Higgs loop 
to which the dilaton couples by the vertex $- (D-4) 3! \zeta \sqrt{\frac{\lambda_H}{2}} m_h$ in $V^{(1)} (\hat H)$ and
with the Higgs self-coupling vertex $- 3! \sqrt{\frac{\lambda_H}{2}} m_h$ in $V^{(0)} (\hat H)$.  
The amplitude ${\cal{T}}_{A3}$ reads
\begin{eqnarray}
{\cal{T}}_{A3} &=& 2 i (D-4) \zeta \left(- 3! \sqrt{\frac{\lambda_H}{2}} m_h \right)^2  \int \frac{d^D k}{(2 \pi)^D} 
\frac{1}{(k^2 + m_h^2)\left[(k+q)^2 + m_h^2 \right]}
\nonumber\\
&=& 36 i \zeta  \lambda_H  m_h^2 (D -4) \frac{i}{16 \pi^2} \Gamma(2 - \frac{D}{2})
\nonumber\\
&=& \frac{9}{2 \pi^2} \zeta  \lambda_H  m_h^2,
\label{T-A3-1}
\end{eqnarray}
where $q$ is the external momentum of the Higgs field. In order to reach the final result in Eq.  (\ref{T-A3-1}), 
we have evaluated the integral as follows: 
\begin{eqnarray}
I &=& \int d^D k \frac{1}{(k^2 + m_h^2)\left[(k+q)^2 + m_h^2 \right]}
\nonumber\\
&=& \int_0^1 d x \int d^D k \frac{1}{\left[ (k^2 + m_h^2) (1-x) + ( (k+q)^2 + m_h^2) x \right]^2}
\nonumber\\
&=& \int_0^1 d x \int d^D k \frac{1}{\left[ (k+ xq)^2 + m_h^2 + x (1-x) q^2 \right]^2}
\nonumber\\
&=& \int_0^1 d x \int d^D k \frac{1}{\left[ k^2 + m_h^2 + x (1-x) q^2 \right]^2}
\nonumber\\
&=& \int_0^1 d x \ i  \pi^2 \Gamma(2-\frac{D}{2}) (m_h^2)^{\frac{D}{2} -2} (1 - x + x^2)^{\frac{D}{2} -2}
\nonumber\\
&=& i \pi^2 \Gamma(2-\frac{D}{2}).
\label{Q-A3-1}
\end{eqnarray}
Here at the second equality, we have used the Feynman parameter formula (\ref{App-B-Feynman2})
and at the fourth equality, we have shifted the momentum $k + x q \rightarrow k$, which is allowed
since the integral is now finite owing to the regularization, and at the fifth equality we have used 
the on-mass-shell condition $q^2 = - m_h^2$ and Eq.  (\ref{Q-F2}).

Thus, adding three types of contributions, we have 
\begin{eqnarray}
{\cal{T}}_A = {\cal{T}}_{A1} + {\cal{T}}_{A2} +  {\cal{T}}_{A3} = \frac{6}{\pi^2} \zeta  \lambda_H  m_h^2.
\label{All-A}
\end{eqnarray}
From this result, we can construct an effective Lagrangian at the one-loop level
\begin{eqnarray}
L_{\sigma h^2} = - \frac{3}{\pi^2} \zeta  \lambda_H  \frac{m_h^2}{M_p} \sigma h^2,
\label{Eff-A}
\end{eqnarray}
where we have explicitly written down the Planck mass dependence in such a way that we can
recognize dimensions clearly. 

Next, let us take account of four-point (with three Higgs and one dilaton as the external
particles), one-loop diagrams. In this case, inspection of the vertices reveals again that 
there are two types of one-loop divergent diagrams where the Higgs field is circulating in the loop.
One type of one-loop divergent diagram, which we call the diagram (B1), is given by the Higgs loop 
to which the dilaton couples by the vertex $- (D-4) 3! \zeta \lambda_H$ in $V^{(1)} (\hat H)$
and with the Higgs self-coupling $- 3! \sqrt{\frac{\lambda_H}{2}} m_h$ in $V^{(0)} (\hat H)$. 
The corresponding amplitude ${\cal{T}}_{B1}$ reads
\begin{eqnarray}
{\cal{T}}_{B1} &=& i (D-4) 3! \zeta \lambda_H \sqrt{\frac{\lambda_H}{2}} m_h 3!  \int \frac{d^D k}{(2 \pi)^D} 
\frac{1}{(k^2 + m_h^2)\left[(k + q)^2 + m_h^2 \right]}
\nonumber\\
&=& 36 i \zeta \lambda_H \sqrt{\frac{\lambda_H}{2}} m_h (D-4) \frac{i}{16 \pi^2} \Gamma(2 - \frac{D}{2})
\nonumber\\
&=& \frac{9}{2 \sqrt{2} \pi^2} \zeta  \lambda_H  \sqrt{\lambda_H} m_h,
\label{T-B1-1}
\end{eqnarray}
where we have used Eq. (\ref{Q-A3-1}).

The other type of one-loop divergent diagram, which is called the diagram (B2), is given by the Higgs loop 
to which the dilaton couples by the vertex $- (D-4) 3! \zeta \sqrt{\frac{\lambda_H}{2}} m_h$ in $V^{(1)} (\hat H)$
and with the Higgs self-coupling $- 3! \lambda_H$ in $V^{(0)} (\hat H)$. 
The amplitude ${\cal{T}}_{B2}$ takes the form
\begin{eqnarray}
{\cal{T}}_{B2} &=& i (D-4) 3! \zeta \sqrt{\frac{\lambda_H}{2}} m_h 3! \lambda_H \int \frac{d^D k}{(2 \pi)^D} 
\frac{1}{(k^2 + m_h^2)\left[(k + q)^2 + m_h^2 \right]}
\nonumber\\
&=& \frac{9}{2 \sqrt{2} \pi^2} \zeta  \lambda_H  \sqrt{\lambda_H} m_h,
\label{T-B2-1}
\end{eqnarray}

Putting the two types of contributions together, we obtain
\begin{eqnarray}
{\cal{T}}_B = {\cal{T}}_{B1} + {\cal{T}}_{B2} = \frac{9}{\sqrt{2} \pi^2} \zeta  \lambda_H \sqrt{\lambda_H} m_h.
\label{All-B}
\end{eqnarray}
This result gives rise to an effective Lagrangian 
\begin{eqnarray}
L_{\sigma h^3} = - \frac{3}{2 \sqrt{2} \pi^2} \zeta  \lambda_H \sqrt{\lambda_H} \frac{m_h}{M_p} \sigma h^3,
\label{Eff-B}
\end{eqnarray}
where we have recovered the Planck mass again.

Now we turn our attention to five-point (with four Higgs and one dilaton as the external
particles), one-loop diagrams. In this case, we find that there is only one type of one-loop divergent diagram 
where the Higgs field is circulating in the loop.
This type of one-loop divergent diagram, which we call the diagram (C), is given by the Higgs loop 
to which the dilaton couples by the vertex $- (D-4) 3! \zeta \lambda_H$ in $V^{(1)} (\hat H)$
and with the Higgs self-coupling $- 3! \lambda_H$ in $V^{(0)} (\hat H)$. 
The corresponding amplitude ${\cal{T}}_{C}$ reads
\begin{eqnarray}
{\cal{T}}_{C} &=& 2 i (D-4) 3! \zeta \lambda_H 3! \lambda_H \int \frac{d^D k}{(2 \pi)^D} 
\frac{1}{(k^2 + m_h^2)\left[(k + p + q)^2 + m_h^2 \right]}
\nonumber\\
&=& \frac{1}{\pi^2} \zeta  \lambda_H^2,
\label{T-C-1}
\end{eqnarray}
where $p$ and $q$ are external momenta of the two Higgs fields. This quantum effect gives us
an effective Lagrangian
\begin{eqnarray}
L_{\sigma h^4} = - \frac{1}{24 \pi^2} \zeta  \lambda_H^2 \frac{1}{M_p} \sigma h^4.
\label{Eff-C}
\end{eqnarray}

Finally, it is straightforward to evaluate $(n+1)$-point (with $n \ge 5$ Higgs and one dilaton as external particles),
one-loop diagrams in a similar manner. It turns out that these higher-point, one-loop diagrams do not
yield any divergence, thereby leading to the vanishing effective Lagrangian. Moreover, we find that 
there are no divergences for the $\sigma^n h^m (n \ge 2)$-type of amplitudes at the one-loop level.

After all, we have a total effective Lagrangian at the one-loop level 
\begin{eqnarray}
L^{1-loop} = \left[ - \frac{3}{\pi^2} \zeta  \lambda_H  \frac{m_h^2}{M_p} h^2
- \frac{3}{2 \sqrt{2} \pi^2} \zeta  \lambda_H \sqrt{\lambda_H} \frac{m_h}{M_p} h^3
- \frac{1}{24 \pi^2} \zeta  \lambda_H^2 \frac{1}{M_p} h^4 \right]  \sigma.
\label{Eff-All}
\end{eqnarray}
Note that this effective Lagrangian has the similar form to the potential as $V^{(1)} (\hat H)$
but each coefficient is suppressed by the Planck mass, which means that effects of radiative
corrections are very tiny in the low energy region.

\subsection{Yukawa coupling}

In the standard model, the fermion masses arise from the Yukawa coupling between the fermions and 
the Higgs field. It is therefore of interest to evaluate radiative corrections of the Yukawa coupling in the
present model.

It is easy to see that there are no radiative corrections to the coupling between the dilaton and 
the fermions at the one-loop level, but it turns out that the one-loop induced vertex produces 
radiative corrections to this coupling, so we are willing to calcuclate this quantum effect in this
subsection. 

Before delving into the calculation, let us go back to the basics of the Yukawa coupling.
The Yukawa coupling between the fermions and the Higgs field is generically given by the following
Lagrangian
\begin{eqnarray}
{\cal{L}}_{H \bar \psi \psi} = - \sqrt{-g}  g_Y \bar \psi_L H \psi_R,
\label{Yukawa1}
\end{eqnarray}
where $g_Y$ is the Yukawa coupling constant, $\psi_L$ and $\psi_R$ are respectively
a left-handed, $SU(2)$-doublet spinor and a right-handed singlet spinor.   

To move the Jordan frame to the Einstein frame, we use the local scale transformation 
(\ref{Q-L-scale transf}) and its fermionic one
\begin{eqnarray}
\hat \psi_L = \Omega^{- \frac{D+2}{4}}(x) \psi_L,  \quad 
\hat \psi_R = \Omega^{- \frac{D+2}{4}}(x) \psi_R.
\label{Q-L-scale transf 2}
\end{eqnarray}
Under this local scale transformation, the Lagrangian (\ref{Yukawa1}) takes the same form
\begin{eqnarray}
{\cal{L}}_{H \bar \psi \psi} = - \sqrt{- \hat g}  g_Y \bar {\hat \psi_L}  \hat H  \hat \psi_R.
\label{Yukawa2}
\end{eqnarray}
With the following definitions of spinors and the unitary gauge for the Higgs field $\hat H$,
\begin{eqnarray}
\hat \psi_L^T = (\hat \chi, \hat \psi),  \quad   
\hat \psi_R = \hat \psi,   \quad
\hat H^T = ( 0, \frac{v + h(x)}{\sqrt{2}} ),
\label{Yukawa3}
\end{eqnarray}
the Lagrangian is reduced to
\begin{eqnarray}
{\cal{L}}_{H \bar \psi \psi} = - \sqrt{- \hat g}  ( M_\psi \bar {\hat \psi} \hat \psi
+ \frac{g_Y}{\sqrt{2}} \bar {\hat \psi}  \hat \psi h ),
\label{Yukawa4}
\end{eqnarray}
where we have defined $M_\psi = \frac{g_Y}{\sqrt{2}} v$.

We are now in a position to calculate the one-loop amplitude where two fermions and one dilaton
appear as the external particles. In this case, there is no divergent diagram but we have a finite diagram 
where the fermion and the Higgs field propagate in the loop, which we call the diagram (D). In this diagram,
the dilaton couples to the Higgs by the vertex $- \frac{6}{\pi^2} \zeta \lambda_H \frac{m_h^2}{M_p}$ 
in (\ref{Eff-All}), which is a one-loop effect\footnote{As will seen later, we also have the similar contribution 
from the $U(1)$ gauge sector at the one-loop level, but we will now neglect it since the contribution from 
the gauge sector is smaller than that from (\ref{Eff-All}).}, and two fermions couple to the Higgs by
the vertex $- \frac{g_Y}{\sqrt{2}}$ in (\ref{Yukawa4}).  Thus, this diagram is essentially a two-loop
effect. The correponding amplitude is given by 
\begin{eqnarray}
{\cal{T}}_D &=& - i \frac{6}{\pi^2} \zeta \lambda_H \frac{m_h^2}{M_p} (\frac{g_Y}{\sqrt{2}})^2 
\int \frac{d^D k}{(2 \pi)^D} 
\frac{\slash{q} - \slash{k} - M_\psi}{(k^2 + m_h^2)^2 \left[(q-k)^2 + M_\psi^2 \right]}
\nonumber\\
&=& \frac{3}{16 \pi^4} \zeta  \lambda_H  g_Y^2 \frac{m_h^2}{M_\psi M_p} f(\frac{M_\psi}{m_h}),
\label{T-D-1}
\end{eqnarray}
where $q$ is the external momentum of the fermion field, which satisfies the on-mass-shell condition
$q^2 = - M_\psi^2$ and the function $f(x)$ is defined as
\begin{eqnarray}
f(x) = \log x + \frac{1 - 2 x^2}{\sqrt{1 - 4 x^2}} \log \left[ \frac{1}{2 x} ( 1 + \sqrt{1 - 4 x^2} ) \right]. 
\label{f(x)}
\end{eqnarray}

To obtain the result in Eq.  (\ref{T-D-1}), we have calculated the integral as follows: 
\begin{eqnarray}
J &=& \int d^D k \frac{\slash{q} - \slash{k} - M_\psi}{(k^2 + m_h^2)^2 \left[(q-k)^2 + M_\psi^2 \right]}
\nonumber\\
&=& \int_0^1 d x \int d^D k \frac{2 (1 - x) (\slash{q} - \slash{k} - M_\psi) }
{\left[ (k^2 + m_h^2) (1-x) + ( (q-k)^2 + M_\psi^2) x \right]^3}
\nonumber\\
&=& \int_0^1 d x \int d^D k \frac{2 (1 - x) (\slash{q} - \slash{k} - M_\psi) }
{\left[ (k- xq)^2 + M_\psi^2 x^2 - m_h^2 x + m_h^2 \right]^3}
\nonumber\\
&=& \int_0^1 d x \int d^D k \frac{2 (1 - x) [ (1 - x ) \slash{q} - \slash{k} - M_\psi) }
{\left[ k^2 + M_\psi^2 x^2 - m_h^2 x + m_h^2 \right]^3}
\nonumber\\
&=& - 2 M_\psi \int_0^1 d x \int d^D k \frac{1 - x}
{\left[ k^2 + M_\psi^2 x^2 - m_h^2 x + m_h^2 \right]^3}
\nonumber\\
&=& \frac{i \pi^2}{M_\psi} f(\frac{M_\psi}{m_h}).
\label{Q-D-2}
\end{eqnarray}
Here at the second equality, we have used the Feynman parameter formula (\ref{App-B-Feynman3}).
At the fourth equality, we have shifted the momentum $k - x q \rightarrow k$, and
used that $\int d^D k \ k^\mu F(k^2) = 0$ for a general function $F$ in addition to $q^\mu \approx 0$
at the low energy.  Furthermore, at the final equality, we have made use of the integral formula
\begin{eqnarray}
\int d^D k \frac{1}{( k^2 + \Delta )^3} = \frac{i \pi^2}{2 \Delta},
\label{Q-D-3}
\end{eqnarray}
which is a specific case of a general formula (\ref{App-B-Integral3}). 

From the above result, an effective Lagrangian for the interaction between the dilaton and fermions 
can be derived to 
\begin{eqnarray}
L_{\sigma \bar \psi \psi} = - g_\sigma \bar{\hat \psi} \hat \psi \sigma,
\label{Q-D-4}
\end{eqnarray}
where the effective coupling $g_\sigma$ is defined by the absolute value of ${\cal{T}}_D$, 
i.e., $g_\sigma = | {\cal{T}}_D |$.

\subsection{Dilaton mass}

As seen in the Lagrangian (\ref{Q-Lagr 4}), the dilaton is exactly massless at the classical
level since it is the Nambu-Goldstone boson stemming from spontaneous symmetry breakdown
of scale symmetry. However, it is well-known that the scale symmetry is violated by
the trace anomaly at the quantum-mechanical level, and as a result, the dilaton becomes
massive. 

It is very interesting to evaluate the size of the dilaton mass within the present formalism.
It is in general expected that if any, the Nambu-Goldstone boson
would not be so heavy as in the pions. Of course, the size of the dilaton mass would be
closely related to an energy scale where the scale symmetry is broken spontaneously.
On the other hand, because the dilaton is a representative example of scalar particle as well as the Higg particle,
it is of interest to investigate if the dilaton would receive the quadratic divergence like the Higgs particle
or not. 
 
It turns out that at the one-loop effect, there is no quantum correction for the self-energy of the dilaton
and it is at the two-loop effect that radiative corrections appear for it in the formalism at hand.
Actually, we have a one-loop divergent diagram for the self-energy of the dilaton, which we call the diagram (E), 
where two external dilatons couple to the Higgs loop by the vertex $- (D-4) \zeta m_h^2$ in $V^{(1)} (\hat H)$ 
and the vertex $-\frac{6}{\pi^2} \zeta \lambda_H \frac{m_h^2}{M_p}$ in (\ref{Eff-All}) which is already a one-loop effect.
Therefore, this one-loop diagram is essentially a two-loop contribution. 
The amplitude takes the form 
\begin{eqnarray}
{\cal{T}}_E &=& 2 i (D-4) \zeta m_h^2 \frac{6}{\pi^2} \zeta \lambda_H \frac{m_h^2}{M_p} 
\int \frac{d^D k}{(2 \pi)^D} \frac{1}{(k^2 + m_h^2)^2}
\nonumber\\
&=& \frac{3}{2 \pi^4} \zeta^2  \lambda_H  \frac{m_h^4}{M_p}.
\label{Q-E-1}
\end{eqnarray}

This amplitude directly gives rise to an effective action for the mass term of the dilaton
\begin{eqnarray}
L_{\sigma^2}^{(2)} = - \frac{3}{4 \pi^4} \zeta^2  \lambda_H  \frac{m_h^4}{M_p^2} \sigma^2
\equiv - \frac{1}{2} m_\sigma^2 \sigma^2,
\label{Q-E-2}
\end{eqnarray}
where we have defined the induced dilaton mass $m_\sigma$ as
\begin{eqnarray}
m_\sigma^2 = \frac{3}{2 \pi^4} \zeta^2  \lambda_H  \frac{m_h^4}{M_p^2}.
\label{Q-E-3}
\end{eqnarray}
As expected, it has turned out that the dilaton, which is massless classically, becomes massive
because of radiative corrections.

From the result  (\ref{Q-E-3}), one might be tempted to conclude that the dilaton mass induced by 
radiative corrections is very small since the size of the mass is suppressed by the Planck mass
and $\zeta \approx \lambda_H \approx {\cal {O}}(1)$.  But the story has not ended yet 
because we have to take the quadratic divergence, which is the root of the hiearchy problem
in case of the Higgs particle, into consideration. Since there is no interaction
vertex $\sigma^2 h^2$ in the present formalism, the most severe quadratic divergence appears when
the fermion is circulating in the loop via the vertex in the Lagrangian (\ref{Q-D-4}). 
The amplitude ${\cal{T}}_F$, which is essentially a five-loop effect, is certainly quadratically divergent 
by power counting
\begin{eqnarray}
{\cal{T}}_F = i g_\sigma^2 \int \frac{d^D k}{(2 \pi)^D} 
\frac{1}{(\slash{k} + M_\psi)^2}
\approx  i g_\sigma^2 \int \frac{d^D k}{(2 \pi)^D} \frac{1}{k^2}
\approx  - g_\sigma^2 \Lambda^2,
\label{T-F-1}
\end{eqnarray}
where $\Lambda$ is the ultra-violet cutoff. Then, with the reasonable choice $\Lambda = M_p$,
the mass of the dilaton is approximately given by
\begin{eqnarray}
m_\sigma \approx g_\sigma \Lambda = \frac{3}{16 \pi^4} \zeta  \lambda_H  g_Y^2 
\frac{m_h^2}{M_\psi} | f(\frac{M_\psi}{m_h}) |,
\label{T-F-2}
\end{eqnarray}
which is around the GeV scale since $| f(\frac{M_\psi}{m_h}) | \approx 1$ for $m_h \approx M_\psi$,
which holds  approximately for $\psi =$ top-quark. (Here it is reasonable to take $\zeta \approx
\lambda_H \approx  g_Y \approx {\cal{O}}(1)$ at the low energy.)
Note that the factor $\frac{1}{M_p}$ in $g_\sigma$ is cancelled by the cutoff $M_p$.
It is remarkable that the quadratic divergence, which leads to the burdensome hierarchy
problem in case of the Higgs particle, gives the GeV scale mass to the dilaton!  At first sight,
it appears that the GeV scale mass of the dilaton is against the results of the LHC owing to
null results in searches for new scalar particles except the Higgs particle below a few TeV scale.
However, as seen in the relation $g_\sigma = | {\cal{T}}_D |$ and Eq. (\ref{T-D-1}), the coupling
between the dilaton and the Higgs particle is so tiny that it is extremely difficult to detect
the dilaton in the LHC.

\subsection{Contributions from gauge fields}

In the previous subsections, we have switched off the $U(1)$ gauge fields. 
In this final subsection, we switch on the $U(1)$ gauge fields, and wish to calculate the coupling between the dilaton and the Higgs
particle by using propagators and vertices from the sector of the gauge fields in the Lagrangian (\ref{Q-Lagr 4}). 
The result is very simple and illuminating in the sense that we can obtain the similar form of the effective Lagrangian
to (\ref{Eff-All}) at the one-loop level, but the coefficient of each term is multiplied by the square of 
the "fine structure constant".

For convenience, let us pick up part of the Lagrangian (\ref{Q-Lagr 4}) which contains 
the gauge fields
\begin{eqnarray}
{\cal L}_{EM} = \sqrt{- \hat g} \ \Big[  -\frac{1}{4} \sum_{i=1}^2 \hat g^{\mu\nu} \hat g^{\rho\sigma} 
\hat F^{(i)}_{\mu\rho} \hat F^{(i)}_{\nu\sigma}
- \frac{\hat e_1^2}{2 \xi_1} \hat g^{\mu\nu} \hat B_\mu^{(1)} \hat B_\nu^{(1)} 
- \frac{\hat e_2^2}{8} \hat g^{\mu\nu} \hat A_\mu^{(2)} \hat A_\nu^{(2)} ( v + h )^2 \Big].
\label{Q-EM-Lagr}
\end{eqnarray}
With the following definitions of the mass of the gauge fields
\begin{eqnarray}
\hat{m}_A^2 = \frac{1}{4} \hat{e}_2^2 v^2,    \quad
\hat{m}_B^2 = \frac{\hat{e}_1^2}{\xi_1}, 
\label{Q-EM-Mass}
\end{eqnarray}
the Lagrangian (\ref{Q-EM-Lagr}) can be rewritten as
\begin{eqnarray}
{\cal L}_{EM} = \sqrt{- \hat g} \ \Big[  -\frac{1}{4} \sum_{i=1}^2 ( \hat F^{(i)}_{\mu\nu} )^2
- \frac{\hat{m}_B^2}{2} ( \hat B_\mu^{(1)} )^2 
- \frac{\hat{m}_A^2}{2} ( \hat A_\mu^{(2)} )^2
-  \hat{m}_A^2 ( \frac{1}{v} h + \frac{1}{2 v^2} h^2 ) ( \hat A_\mu^{(2)} )^2 \Big].
\label{Q-EM-Lagr2}
\end{eqnarray}
Because of the relations $\hat{e}_i = \Omega^{\frac{D-4}{2}} e_i = e^{\frac{D-4}{D-2} \zeta \sigma} e_i \ (i = 1, 2)$,
we have
\begin{eqnarray}
\hat{m}_A^2 \approx m_A^2 [ 1 + (D-4) \zeta \sigma ],    \quad
\hat{m}_B^2 \approx m_B^2 [ 1 + (D-4) \zeta \sigma ], 
\label{Q-EM-Mass2}
\end{eqnarray}
where $m_A^2, m_B^2$ are defined as in (\ref{Q-EM-Mass}) but without the hat on $e_i$.
Thus, in the sector of the gauge fields, we have six different vertices $\sigma-B^2, \sigma-A^2,
h-A^2, h^2-A^2, \sigma-h-A^2, \sigma-h^2-A^2$, and two propagators of massive gauge
fields $\hat A_\mu^{(2)}, \hat B_\mu^{(1)}$ for which we take the Feynman gauge.

Now, on the basis of these vertices and propagators, we would like to consider $(n+1)$-point 
(with $n$ Higgs particles and one dilaton as external particles), one-loop diagrams. 
For $n=2$, we have two types of one-loop divergent diagrams in which the gauge field $A_\mu^{(2)}$ 
is circulating in the loop. One type of the diagram, which we call the diagram (G1), is a tadpole type 
in which the dilaton couples to the loop composed of the gauge field $A_\mu^{(2)}$ 
by the vertex $- 2 \frac{m_A^2}{v^2} (D-4) \zeta \eta_{\mu\nu} = - \frac{1}{2} e_2^2 (D-4) \zeta \eta_{\mu\nu}$.
The corresponding amplitude ${\cal{T}}_{G1}$ is given by  
\begin{eqnarray}
{\cal{T}}_{G1} &=& - 2i e_2^2 (D-4) \zeta  
\int \frac{d^D k}{(2 \pi)^D} \frac{1}{k^2 + m_A^2}
\nonumber\\
&=& \frac{1}{4 \pi^2} e_2^2 \zeta m_A^2
\nonumber\\
&=& \alpha_2^2 \zeta v^2,
\label{Q-G-1}
\end{eqnarray}
where we have introduced the "fine structure constant" $\alpha_2 = \frac{e_2^2}{4 \pi}$.

The other type of one-loop diveregent diagram which we call the diagram (G2), is the self-energy
type of the dilaton where the dilaton couples to the loop by the vertex $- \frac{1}{4} e_2^2 v
(D-4) \zeta \eta_{\mu\nu}$ and with the gauge-Higgs coupling vertex $- \frac{1}{4} e_2^2 v \eta_{\mu\nu}$.
The amplitude ${\cal{T}}_{G2}$ is given by  
\begin{eqnarray}
{\cal{T}}_{G2} &=& i \frac{1}{2}  (D-4) \zeta (e_2^2 v)^2  
\int \frac{d^D k}{(2 \pi)^D} \frac{1}{(k^2 + m_A^2)[ (k+q)^2 + m_A^2 ]}
\nonumber\\
&=& \alpha_2^2 \zeta v^2,
\label{Q-G-2}
\end{eqnarray}
where $q$ is the external momentum of the Higgs particle. 

Hence, adding the two results, we have
\begin{eqnarray}
{\cal{T}}_G = {\cal{T}}_{G1} + {\cal{T}}_{G2} = 2 \alpha_2^2 \zeta v^2,
\label{Q-G}
\end{eqnarray}
from which, we obtain an effective Lagrangian
\begin{eqnarray}
L_{\sigma h^2} = - \alpha_2^2 \zeta \frac{v^2}{M_p} \sigma h^2.
\label{Q-G-L}
\end{eqnarray}

Next, let us move to the evaluation of $4$-point (with $3$ Higgs particles and one dilaton 
as external particles), one-loop diagrams. In this case, there are two kinds of divergent
diagrams. The one diagram, which we call the diagram (H1), has the $h-A^2$ vertex 
with the coefficient $- \frac{1}{4} e_2^2 v \eta_{\mu\nu}$ and $\sigma-h^2-A^2$ vertex 
with the coefficient $- \frac{1}{2} e_2^2 (D-4) \zeta \eta_{\mu\nu}$, so
the amplitude takes the form
\begin{eqnarray}
{\cal{T}}_{H1} &=& i \frac{1}{2} (D-4) \zeta e_2^4 v  
\int \frac{d^D k}{(2 \pi)^D} \frac{1}{(k^2 + m_A^2)[ (k+q)^2 + m_A^2 ]}
\nonumber\\
&=& \alpha_2^2 \zeta v,
\label{Q-H-1}
\end{eqnarray}
where $q$ is the momentum carried by the external Higgs particle coupled to the loop
via the vertex $h-A^2$.

The other divergent diagram, which is called (H2), has the vertex $h^2-A^2$ with
the coefficient $- \frac{1}{2} e_2^2 \eta_{\mu\nu}$ and the vertex $\sigma-h-A^2$
with the coefficient $- \frac{1}{4} e_2^2 v (D-4) \zeta \eta_{\mu\nu}$.
It turns out the corresponding amplitude is the same as ${\cal{T}}_{H1}$, so we have
\begin{eqnarray}
{\cal{T}}_{H2} = \alpha_2^2 \zeta v.
\label{Q-H-2}
\end{eqnarray}
Putting Eqs.  (\ref{Q-H-1}) and (\ref{Q-H-2}) together, we obtain the result
\begin{eqnarray}
{\cal{T}}_H = {\cal{T}}_{H1} + {\cal{T}}_{H2} = 2 \alpha_2^4 \zeta v,
\label{Q-H-3}
\end{eqnarray}
from which an effective Lagrangian becomes
\begin{eqnarray}
L_{\sigma h^3} = - \frac{1}{3} \alpha_2^2 \zeta \frac{v}{M_p} \sigma h^3.
\label{Q-H-L}
\end{eqnarray}

Finally, as $5$-point (with $4$ Higgs particles and one dilaton as external particles), one-loop 
divergent diagrams, there is only one diagram constructed out of the vertex $h^2-A^2$ and the
vertex $\sigma-h^2-A^2$. The amplitude ${\cal{T}}_{G3}$ is of form
\begin{eqnarray}
{\cal{T}}_I &=& 2 i (D-4)  \zeta e_2^4  
\int \frac{d^D k}{(2 \pi)^D} \frac{1}{(k^2 + m_A^2)[ (k+p-q)^2 + m_A^2 ]}
\nonumber\\
&=& 4 \alpha_2^2 \zeta,
\label{Q-I-1}
\end{eqnarray}
from which an effective Lagrangian can be derived to
\begin{eqnarray}
L_{\sigma h^4} = - \frac{1}{6} \alpha_2^2 \zeta \frac{1}{M_p} \sigma h^4.
\label{Q-I-2}
\end{eqnarray}
Incidentally, it is easy to check that there are no divergent, one-loop diagrams for
the case of $(n+1)$-point (with $n \ge 5$ Higgs and one dilaton as external particles) as before.
 
Putting these results together, we have the following effective Lagrangian from the sector
of the gauge fields:
\begin{eqnarray}
L_{EM}^{1-loop} &=& \alpha_2^2 \left[ - \zeta \frac{v^2}{M_p} h^2
- \frac{1}{3} \zeta \frac{v}{M_p} h^3
- \frac{1}{6} \zeta \frac{1}{M_p} h^4 \right]  \sigma
\nonumber\\
&=& \left(\frac{\alpha_2}{\lambda_H} \right)^2 \left[ - \frac{1}{2} \zeta \lambda_H \frac{m_h^2}{M_p} h^2
- \frac{1}{3 \sqrt{2}} \zeta \lambda_H \sqrt{\lambda_H} \frac{m_h}{M_p} h^3
- \frac{1}{6} \zeta \lambda_H^2 \frac{1}{M_p} h^4 \right]  \sigma,
\label{Eff-All2}
\end{eqnarray}
where we have used the relation $v = \frac{m_h}{\sqrt{2 \lambda_H}}$ in (\ref{Q-P3}).
The last equality shows that the Lagrangian (\ref{Eff-All}) is more dominant than the Lagrangian (\ref{Eff-All2}) 
because of $\lambda_H \gg \alpha_2$.
Thus, the results about the dilaton mass obtained in the previous subsection in essence remain
unchanged even if the contribution from the gauge fields is taken into consideration.

\section{Conclusion}

In this article, we have investigated a Higgs mechanism in scale-invariant theories of gravitation 
in detail. After reviewing this new Higgs mechanism found in our previous articles \cite{Oda1, Oda2}
in terms of the simplest model, we have extended the Higgs mechanism to non-Abelian gauge groups
and a scalar field with many components. Since we have already considered the Higgs mechanism
in a locally scale-invariant theory of gravitation, i.e., conformal gravity, the validity of this
mechanism in the scale-invariant gravitational theories is very universal and therefore would have 
some phenomenological applications in future.  

Moreover, we have spelled out quantum effects of a scale-invariant extension of the standard model
in a flat Minkowski background, and examined the coupling between the dilaton and the Higgs particle.
An intriguing observation done in our analysis is that although the mass of the dilaton is exactly
zero at the classical level owing to the Nambu-Goldstone theorem, it becomes non-zero and takes a finite size
around the GeV scale because of radiative corrections. It is worthwhile to mention that we have succeeded 
in deriving the size of the dilaton mass deductively by starting with a fundamental theory and without 
any specific assumption. 
As long as we know, the dilaton mass has not thus far been obtained in such a priori manner,
so we think our derivation of the dilaton mass to be very interesting. As mentioned in the article, the
GeV scale mass of the dilaton is consistent with the recent null results of new scalar particles except the
Higgs particle in the LHC since the coupling constant of the dilaton is too small to detect the dilaton in the
LHC. However, the dilaton with the GeV scale mass would have some implication in cosmology, e.g., 
the dilaton could become one of candidates of dark matter if it is somehow stable because of some
unknown mechanism.  

Our consideration in this article is confined to the quantum analysis in a fixed Minkowski background. In other
words, quantum effects coming from the gravity are completely ignored because of non-renormalizability 
of quantum gravity. Since quantum gravity effects are of course not so dominant as quantum effects from matter fields 
in the low energy region, it is physically reasonable to neglect quantum effects of the gravity as the first approximation
of the calculation. Nevertheless, it is of interest to take into consideration the quantum effects from the gravity.
In the future, we wish to study the quantum effects from the gravitational sector in the present formalism. 

Another interesting study for an application of our finding is the Higgs inflation \cite{Bezrukov}. We wish to
return this problem as well in near future.

\begin{flushleft}
{\bf Acknowledgements}
\end{flushleft}
This work is supported in part by the Grant-in-Aid for Scientific 
Research (C) Nos. 22540287 and 25400262 from the Japan Ministry of Education, Culture, 
Sports, Science and Technology.


\appendix
\section{Derivation of current for scale transformation} \label{App:AppendixA}

In Appendix A, we will present a derivation of the dilatation current  (\ref{Current}) via the Noether theorem.
It is easy to show that the Lagrangian  (\ref{Lagr 1}) is invariant under the scale transformation  (\ref{Scale transf})
without surface terms. Therefore, the expression of the Noether current is of form
\begin{eqnarray}
\Lambda J^\mu = \sum \frac{\partial {\cal L}}{\partial \partial_\mu \phi} \delta \phi,
\label{App-A-Current}
\end{eqnarray}
where $\phi = \{g_{\mu\nu}, \Phi, \Phi^\dagger \}$. Under the scale transformation (\ref{Scale transf})
with a global parameter $\Omega = e^\Lambda \approx 1 + \Lambda \ (|\Lambda| \ll 1)$, the current
reads
\begin{eqnarray}
J^\mu = \frac{\partial {\cal L}}{\partial \partial_\mu g_{\rho\sigma}}  2 g_{\rho\sigma}
- \frac{\partial {\cal L}}{\partial \partial_\mu \Phi} \Phi 
- \frac{\partial {\cal L}}{\partial \partial_\mu \Phi^\dagger} \Phi^\dagger,
\label{App-A-J}
\end{eqnarray}
so we have to calculate three objects $\frac{\partial {\cal L}}{\partial \partial_\mu g_{\rho\sigma}},
\frac{\partial {\cal L}}{\partial \partial_\mu \Phi}, \frac{\partial {\cal L}}{\partial \partial_\mu \Phi^\dagger}$
to obtain the expression of the dilatation current $J^\mu$. In particular, calculating the first object
$\frac{\partial {\cal L}}{\partial \partial_\mu g_{\rho\sigma}}$ is so complicated that we will present 
its derivation in detail.

First, with the definition $\varphi = \xi \Phi^\dagger \Phi$, let us consider the non-minimal term
\begin{eqnarray}
{\cal L}_{NM} = \sqrt{-g} \ \varphi R = {\cal L}_1 +  {\cal L}_2,
\label{App-A-NM}
\end{eqnarray}
where we have defined 
\begin{eqnarray}
{\cal L}_1 &=& \sqrt{-g} \ \varphi g^{\mu\nu} ( \partial_\alpha \Gamma^\alpha_{\mu\nu}
-  \partial_\nu \Gamma^\alpha_{\mu\alpha} ),   \nonumber\\
{\cal L}_2 &=& \sqrt{-g} \ \varphi g^{\mu\nu} ( \Gamma^\alpha_{\sigma\alpha} \Gamma^\sigma_{\mu\nu}
-  \Gamma^\alpha_{\sigma\nu} \Gamma^\sigma_{\mu\alpha} ),
\label{App-A-NM2}
\end{eqnarray}
where as usual the affine connection and its contraction are defined as
\begin{eqnarray}
\Gamma^\alpha_{\mu\nu} = \frac{1}{2} g^{\alpha\beta} ( \partial_\mu g_{\beta\nu} + \partial_\nu g_{\beta\mu}
-  \partial_\beta g_{\mu\nu} ),   \quad 
\Gamma^\alpha_{\mu\alpha} = \frac{1}{2} g^{\alpha\beta} \partial_\mu g_{\alpha\beta}
= \frac{\partial_\mu \sqrt{-g}} {\sqrt{-g}}.
\label{App-A-Affine}
\end{eqnarray}

${\cal L}_1$ includes terms with second derivative of the metric, i.e., $\partial^2 g$, so we need to perform
the integration by parts to transform them to terms with first derivative, i.e., $\partial g$.
After the integration by parts, ${\cal L}_1$ is devided in two parts, one of which contains terms 
proportional to $\partial \varphi$ and the other part does terms proportional to $\varphi$ itself
\begin{eqnarray}
{\cal L}_1 = - \sqrt{-g} \ \partial_\alpha \varphi ( g^{\mu\nu} \Gamma^\alpha_{\mu\nu}
-  g^{\alpha\mu} \Gamma^\beta_{\mu\beta} ) - \varphi [ \partial_\alpha ( \sqrt{-g} g^{\mu\nu} ) \Gamma^\alpha_{\mu\nu} 
- \partial_\nu ( \sqrt{-g} g^{\mu\nu} ) \Gamma^\alpha_{\mu\alpha} ]. 
\label{App-A-L1-1}
\end{eqnarray}
Now let us focus on the second term, which we call $A$, and show that $A$ is equal to $-2 {\cal L}_2$. 
\begin{eqnarray}
A &\equiv& - \varphi [ \partial_\alpha ( \sqrt{-g} g^{\mu\nu} ) \Gamma^\alpha_{\mu\nu} 
- \partial_\nu ( \sqrt{-g} g^{\mu\nu} ) \Gamma^\alpha_{\mu\alpha} ]    \nonumber\\
&=& - \varphi [  \sqrt{-g} ( \Gamma^\beta_{\alpha\beta} g^{\mu\nu} + \partial_\alpha g^{\mu\nu} ) \Gamma^\alpha_{\mu\nu}
- \sqrt{-g} ( \Gamma^\beta_{\nu\beta} g^{\mu\nu} + \partial_\nu g^{\mu\nu} ) \Gamma^\alpha_{\mu\alpha} ].
\label{App-A-L1-2}
\end{eqnarray}
In terms of the definition of the affine connection  (\ref{App-A-Affine}), we can prove the following
relations:
\begin{eqnarray}
\partial_\alpha g^{\mu\nu} \Gamma^\alpha_{\mu\nu} &=& -2 g^{\mu\nu} \Gamma^\alpha_{\sigma\nu} 
\Gamma^\sigma_{\mu\alpha},  \nonumber\\
\partial_\nu g^{\mu\nu} &=& - g^{\alpha\beta} \Gamma^\mu_{\alpha\beta} 
- g^{\mu\nu} \Gamma^\alpha_{\nu\alpha}. 
\label{App-A-L1-3}
\end{eqnarray}
Inserting Eq.  (\ref{App-A-L1-3}) to Eq.  (\ref{App-A-L1-2}), we reach the result that $A$ is equal to
$-2 {\cal L}_2$:
\begin{eqnarray}
A = - 2 \varphi \ \sqrt{-g} g^{\mu\nu} ( \Gamma^\alpha_{\sigma\alpha} \Gamma^\sigma_{\mu\nu}
-  \Gamma^\alpha_{\sigma\nu} \Gamma^\sigma_{\mu\alpha} ) = -2 {\cal L}_2.
\label{App-A-L1-4}
\end{eqnarray}

Next, plugging this result into Eq.  (\ref{App-A-L1-1}) leads to
\begin{eqnarray}
{\cal L}_1 = - \sqrt{-g} \ \partial_\alpha \varphi ( g^{\mu\nu} \Gamma^\alpha_{\mu\nu}
-  g^{\alpha\mu} \Gamma^\beta_{\mu\beta} ) -2 {\cal L}_2.
\label{App-A-L1-5}
\end{eqnarray}
Moreover, substituting Eq.  (\ref{App-A-L1-5}) into Eq.  (\ref{App-A-NM}), we have
\begin{eqnarray}
{\cal L}_{NM} &=& - \sqrt{-g} \ \partial_\alpha \varphi ( g^{\mu\nu} \Gamma^\alpha_{\mu\nu}
-  g^{\alpha\mu} \Gamma^\beta_{\mu\beta} ) - {\cal L}_2    \nonumber\\
&\equiv& {\cal L}_K - {\cal L}_2.
\label{App-A-L1-6}
\end{eqnarray}
Here we have defined
\begin{eqnarray}
{\cal L}_K &=& - \sqrt{-g} \ \partial_\alpha \varphi ( g^{\mu\nu} \Gamma^\alpha_{\mu\nu}
-  g^{\alpha\mu} \Gamma^\beta_{\mu\beta} )    \nonumber\\
&\equiv& - \sqrt{-g} \ \partial_\alpha \varphi K^\alpha,
\label{App-A-K1}
\end{eqnarray}
where $K^\alpha$ is defined as
\begin{eqnarray}
K^\alpha &=& g^{\mu\nu} \Gamma^\alpha_{\mu\nu}
-  g^{\alpha\mu} \Gamma^\beta_{\mu\beta}    \nonumber\\
&=& ( g^{\alpha\rho} g^{\mu\sigma} -  g^{\alpha\mu} g^{\rho\sigma} ) \partial_\mu g_{\rho\sigma},
\label{App-A-K2}
\end{eqnarray}
where at the second equality we have used Eqs. (\ref{App-A-Affine}) and (\ref{App-A-L1-3}).
With this expression  (\ref{App-A-K2}), it is straightforward to take the variation of ${\cal L}_K$ 
with respect to $\partial_\mu g_{\rho\sigma}$ whose result is given by
\begin{eqnarray}
\frac{\partial {\cal L}_K}{\partial \partial_\mu g_{\rho\sigma}}  
= - \sqrt{-g} \ \partial_\alpha \varphi ( g^{\alpha(\rho} g^{\sigma)\mu} -  g^{\alpha\mu} g^{\rho\sigma} ).
\label{App-A-K3}
\end{eqnarray}
Thus, we have 
\begin{eqnarray}
\frac{\partial {\cal L}_K}{\partial \partial_\mu g_{\rho\sigma}}  2 g_{\rho\sigma}   
= 6 \sqrt{-g} \ \partial^\mu \varphi.
\label{App-A-K4}
\end{eqnarray}

Taking the variation of ${\cal L}_2$ with respect to $\partial_\mu g_{\rho\sigma}$ is a bit tedious but
straightforward since the whole calculation can be performed by using the formula
\begin{eqnarray}
\frac{\partial \Gamma^\lambda_{\alpha\beta}}{\partial \partial_\mu g_{\rho\sigma}}   
= \frac{1}{2} [ g^{\lambda\rho} \delta^\mu_{(\alpha} \delta^\sigma_{\beta)}
+ g^{\lambda\sigma} \delta^\mu_{(\alpha} \delta^\rho_{\beta)}
- g^{\lambda\mu} \delta^\rho_{(\alpha} \delta^\sigma_{\beta)} ].
\label{App-A-L2-1}
\end{eqnarray}
After a straightforward calculation using Eq. (\ref{App-A-L2-1}), we have the result
\begin{eqnarray}
\frac{\partial {\cal L}_2}{\partial \partial_\mu g_{\rho\sigma}}  
= \sqrt{-g} \ \varphi [ \frac{1}{2} g^{\rho\sigma} g^{\alpha\beta} \Gamma^\mu_{\alpha\beta} 
-  g^{\rho\alpha} g^{\sigma\beta} \Gamma^\mu_{\alpha\beta} 
+ \frac{1}{2}  ( g^{\mu\rho} g^{\nu\sigma} +  g^{\mu\sigma} g^{\nu\rho} 
- g^{\mu\nu} g^{\rho\sigma} ) \Gamma^\alpha_{\nu\alpha} ].
\label{App-A-L2-2}
\end{eqnarray}
Hence, we obtain
\begin{eqnarray}
\frac{\partial {\cal L}_2}{\partial \partial_\mu g_{\rho\sigma}}  2 g_{\rho\sigma}  
= 2 \sqrt{-g} \ \varphi K^\mu.
\label{App-A-L2-3}
\end{eqnarray}
Accordingly, Eqs.  (\ref{App-A-K4}) and  (\ref{App-A-L2-3}) give us 
\begin{eqnarray}
\frac{\partial {\cal L}_{NM}}{\partial \partial_\mu g_{\rho\sigma}}  2 g_{\rho\sigma}  
= 2 \sqrt{-g} ( 3 \partial^\mu \varphi - \varphi K^\mu).
\label{App-A-NM-2}
\end{eqnarray}
Since $\partial_\mu g_{\rho\sigma}$ is only included in $ {\cal L}_{NM}$, from the
definition $\varphi = \xi \Phi^\dagger \Phi$, we have
\begin{eqnarray}
\frac{\partial {\cal L}}{\partial \partial_\mu g_{\rho\sigma}}  2 g_{\rho\sigma}  
= 2 \xi \sqrt{-g} [ 3 g^{\mu\nu} \partial_\nu (\Phi^\dagger \Phi)  - \Phi^\dagger \Phi K^\mu ].
\label{App-A-g}
\end{eqnarray}

Furthermore, it is easy to calculate the variation of the Lagrangian with respect to $\partial_\mu \Phi,
\partial_\mu \Phi^\dagger$. The results read
\begin{eqnarray}
\frac{\partial {\cal L}}{\partial \partial_\mu \Phi}  \Phi  
&=& - \sqrt{-g} [ \xi \Phi^\dagger \Phi K^\mu + g^{\mu\nu}  (D_\nu \Phi)^\dagger \Phi ],
\nonumber\\
\frac{\partial {\cal L}}{\partial \partial_\mu \Phi^\dagger}  \Phi^\dagger  
&=& - \sqrt{-g} [ \xi \Phi^\dagger \Phi K^\mu + g^{\mu\nu}  (D_\nu \Phi) \Phi^\dagger ].
\label{App-A-Phi}
\end{eqnarray}

Putting together Eqs. (\ref{App-A-g}) and (\ref{App-A-Phi}), the dilatation current (\ref{App-A-J}) is
calculated to be 
\begin{eqnarray}
J^\mu =  ( 6 \xi + 1 ) \sqrt{-g} g^{\mu\nu} \partial_\nu \left(\Phi^\dagger \Phi \right).
\label{App-A-Current2}
\end{eqnarray}

\section{Useful formulae in the loop calculation} \label{App:AppendixB}
 
In Appendix B, we summarize useful formulae in calculating radiative corrections in
Section 5. Following Ref. \cite{Fujii1}, as a regularization method, we adopt the method
of continuous space-time dimensions $D$ in a flat Minkowski space-time. In this regularization method,
all the quantities are extended from four dimensions to $D$ dimensions. Let us therefore focus
on the following loop integral:
\begin{eqnarray}
I (m, n) =  \int d^D k \frac{(k^2)^{m-2}}{(k^2 + \Delta)^n},
\label{App-B-Integral1}
\end{eqnarray}
where $m, n$ are integers and $\Delta$ is a constant. By power counting, this integral is convergent
as long as $D < 2n - 2m + 4$. 

With a Wick rotation $k^0 = i k^D$ and the ansatz of spherical symmetry, the integral  (\ref{App-B-Integral1})
can be rewritten as
\begin{eqnarray}
I (m, n) =  i V(D) \int_0^\infty d k \frac{k^{2m + D - 5}}{(k^2 + \Delta)^n},
\label{App-B-Integral2}
\end{eqnarray}
where $V(D) = \frac{2 \pi^{\frac{D}{2} } } {\Gamma(\frac{D}{2})}$ is a D-dimensional volume form, e.g., 
$V(4) = 2 \pi^2$. Via the change of variables from $k$ to $t = \frac{k^2}{\Delta}$, the integral is
reduced to
\begin{eqnarray}
I (m, n) &=&  i V(D) \frac{1}{2} \Delta^{m-n+\frac{D}{2} -2} \int_0^\infty d t  \frac{t^{m + \frac{D}{2} - 3}}{(1 + t)^n}
\nonumber\\
&=&  i V(D) \frac{1}{2} \Delta^{m-n+\frac{D}{2} -2} B(m + \frac{D}{2} -2, n - m - \frac{D}{2} + 2),
\label{App-B-Integral3}
\end{eqnarray}
where the definition of the beta function is used:
\begin{eqnarray}
B(\alpha, \beta) = \int_0^1 d x \ x^{\alpha - 1} (1 - x)^{\beta-1} 
= \int_0^\infty d t \ t^{\alpha - 1} (1 + t)^{-\alpha-\beta}
= \frac{\Gamma(\alpha) \Gamma(\beta)}{\Gamma(\alpha + \beta)}.
\label{App-B-Beta}
\end{eqnarray}
Since $\Gamma(z)$ has isolated poles at $z = 0, -1, -2, \cdots$, the integral (\ref{App-B-Integral3})
has isolated poles at $D = 2 (n - m + 2), 2 (n - m + 3), \cdots$. We often make use of a relation 
for the gamma function, which holds for positive real numbers $x > 0$
\begin{eqnarray}
\Gamma (x+1) = x \Gamma(x),
\label{App-B-Gamma}
\end{eqnarray}
and $\Gamma(1) = 1$.

In Section 5, to combine propagator denominators we utilize the Feyman parameter formula
\begin{eqnarray}
\frac{1}{A_1 A_2 \cdots A_n} = \int_0^1 d x_1 \cdots d x_n \delta(\sum x_i -1) 
\frac{( n - 1 )! } {(x_1 A_1 + x_2 A_2 + \cdots + x_n A_n)^n}.
\label{App-B-Feynman1}
\end{eqnarray}
In the case of only two denominator factors, this formula reduces to
\begin{eqnarray}
\frac{1}{A B} = \int_0^1 d x  \frac{1} {[x A + (1-x) B]^2}.
\label{App-B-Feynman2}
\end{eqnarray}
Taking differentiation of Eq.  (\ref{App-B-Feynman2}) with respect to $B$, we can derive
another formula
\begin{eqnarray}
\frac{1}{A B^2} = \int_0^1 d x  \frac{2 (1-x)} {[x A + (1-x) B]^3}.
\label{App-B-Feynman3}
\end{eqnarray}


\end{document}